\documentclass[11pt,a4paper,useAMS,apj]{emulateapj}

\usepackage{tikz}
\usetikzlibrary{arrows,shapes,backgrounds}

\usepackage{xcolor}
\usepackage{threeparttable}
\usepackage{multirow}
\usepackage{amsmath}
\usepackage{graphicx}
\usepackage{morefloats}
\usepackage{natbib}
\usepackage[percent]{overpic}
\usepackage{array}
\usepackage[export]{adjustbox}
\usepackage{float}

\definecolor{red}{HTML}{CC0000}
\definecolor{green}{HTML}{007A00}

\slugcomment{LLNL-JRNL-676960}

\shorttitle{The Chandra View of MACS~J1149.6+2223}
\shortauthors{Ogrean et al.}

\newcommand{\chandra}{\emph{Chandra}}
\newcommand{\rosat}{\emph{ROSAT}}

\newcommand{\xmm}{\emph{XMM-Newton}}

\begin{document}


\title{Frontier Fields Clusters: Deep Chandra Observations \\ of the Complex Merger MACS~J1149.6+2223}


\author{
G.~A.~Ogrean\altaffilmark{1,$\star$}, R.~J.~van Weeren\altaffilmark{1,$\dagger$}, C.~Jones\altaffilmark{1}, W.~Forman\altaffilmark{1}, W.~A.~Dawson\altaffilmark{2}, N.~Golovich\altaffilmark{3}, F.~Andrade-Santos\altaffilmark{1}, \and S.~S.~Murray\altaffilmark{1,4}, P.~Nulsen\altaffilmark{1,5}, E.~Roediger\altaffilmark{6,7}, A.~Zitrin\altaffilmark{8,$\star$}, E.~Bulbul\altaffilmark{1}, R.~Kraft\altaffilmark{1}, A.~Goulding\altaffilmark{9},  K.~Umetsu\altaffilmark{10}, T.~Mroczkowski\altaffilmark{11,$\ddagger$}, A.~Bonafede\altaffilmark{12}, S.~Randall\altaffilmark{1}, J.~Sayers\altaffilmark{8}, E.~Churazov\altaffilmark{13,14}, L.~David\altaffilmark{1}, J.~Merten\altaffilmark{15}, M.~Donahue\altaffilmark{16}, B.~Mason\altaffilmark{17}, P.~Rosati\altaffilmark{18}, A.~Vikhlinin\altaffilmark{1}, H.~Ebeling\altaffilmark{19}
}

\affil{\altaffilmark{}}
\affil{\altaffilmark{1}Harvard-Smithsonian Center for Astrophysics, 60 Garden Street, Cambridge, MA 02138, USA; \href{mailto:gogrean@cfa.harvard.edu}{gogrean@cfa.harvard.edu}} 
\affil{\altaffilmark{2}Lawrence Livermore National Lab, 7000 East Avenue, Livermore, CA 94550, USA;}
\affil{\altaffilmark{3}University of California, One Shields Avenue, Davis, CA 95616, USA;}
\affil{\altaffilmark{4}Department of Physics and Astronomy, Johns Hopkins University, 3400 N. Charles Street, Baltimore, MD 21218, USA;}
\affil{\altaffilmark{5}ICRAR, University of Western Australia, 35 Stirling Hwy, Crawley WA 6009, Australia;}
\affil{\altaffilmark{6}Astronomy and Astrophysics Section, Dublin Institute for Advanced Studies, 31 Fitzwilliam Place, Dublin 2, Ireland;}
\affil{\altaffilmark{7}E.~A.~Milne Centre for Astrophysics, Department of Physics and Mathematics, University of Hull, Hull, HU6~7RX, UK;}
\affil{\altaffilmark{8}Cahill Center for Astronomy and Astrophysics, California Institute of Technology, MC 249-17, Pasadena, CA 91125, USA;}
\affil{\altaffilmark{9}Department of Astrophysical Sciences, Princeton University, Princeton, NJ 08544, USA;}
\affil{\altaffilmark{10}Institute of Astronomy and Astrophysics, Academia Sinica, PO Box 23-141, Taipei 10617, Taiwan;}
\affil{\altaffilmark{11}U.S. Naval Research Laboratory, 4555 Overlook Ave SW, Washington, DC 20375, USA;}
\affil{\altaffilmark{12}Hamburger Sternwarte, Universit{\''a}t Hamburg, Gojenbergsweg 112, 21029 Hamburg, Germany;}
\affil{\altaffilmark{13}Max Planck Institute for Astrophysics, Karl-Schwarzschild-Str. 1, 85741, Garching, Germany;}
\affil{\altaffilmark{14}Space Research Institute, Profsoyuznaya 84/32, Moscow, 117997, Russia;}
\affil{\altaffilmark{15}Department of Physics, University of Oxford, Keble Road, Oxford OX1 3RH, UK;}
\affil{\altaffilmark{16}Department of Physics and Astronomy, Michigan State University, East Lansing, MI 48824, USA;}
\affil{\altaffilmark{17}National Radio Astronomy Observatory, 520 Edgemont Road, Charlottesville, VA 22903, USA;}
\affil{\altaffilmark{18}Department of Physics and Earth Science, University of Ferrara, Via G. Saragat, 1-44122 Ferrara, Italy}
\affil{\altaffilmark{19}Institute for Astronomy, University of Hawaii, 2680 Woodlawn Drive, Honolulu, HI 96822, USA;}
\affil{\emph{Submitted to ApJ. Draft version dated \today.}}


\altaffiltext{$\star$}{Hubble Fellow}
\altaffiltext{$\dagger$}{Einstein Fellow}
\altaffiltext{$\ddagger$}{National Research Council Fellow}
\altaffiltext{}{}


\begin{abstract}
    \noindent The HST Frontier Fields cluster MACS~J1149.6+2223 is one of the most complex merging clusters, believed to consist of four dark matter halos. We present results from deep (365~ks) \chandra\ observations of the cluster, which reveal the most distant cold front ($z=0.544$) discovered to date. In the cluster outskirts, we also detect hints of a surface brightness edge that could be the bow shock preceding the cold front. The substructure analysis of the cluster identified several components with large relative radial velocities, thus indicating that at least some collisions occur almost along the line of sight. The inclination of the mergers with respect to the plane of the sky poses significant observational challenges at X-ray wavelengths. MACS~J1149.6+2223 possibly hosts a steep-spectrum radio halo. If the steepness of the radio halo is confirmed, then the radio spectrum, combined with the relatively regular ICM morphology, could indicate that MACS~J1149.6+2223 is an old merging cluster.

\end{abstract}


\keywords{Galaxies: clusters: individual: MACS J1149.6+2223 --- Galaxies: clusters: intracluster medium --- X-rays: galaxies: clusters}




\section{Introduction}

Galaxy clusters grow hierarchically by mergers with other clusters and groups of galaxies, and by accretion of uncollapsed gas and dark matter from the intergalactic medium. Signs of this growth process include features of ram-pressure stripping and instabilities \citep[e.g.,][]{Nulsen1982}, shocks and cold fronts \citep[e.g.,][]{Markevitch2007}, diffuse radio emission \citep[e.g.,][]{Feretti2012}, and significant offsets between the gas and dark matter substructures \citep[e.g.,][]{Clowe2004}.

MACS~J1149.6+2223 is a hot ($T_{\rm 1000}=9.1\pm 0.7$~keV)\footnote{$T_{\rm 1000}$ refers to the temperature measured in a circle of radius $R_{\rm 1000}$ around the cluster center, where $R_{\rm 1000}$ is the radius within which the mean mass density is 1000 times the critical density of the Universe at the cluster redshift. For MACS~J1149.6+2223, $R_{\rm 1000} = 0.83$~Mpc ($R_{\rm 1000} \sim 0.45\,R_{\rm 200}$).} galaxy cluster at $z=0.544$ \citep{Ebeling2007}. The cluster was discovered as part of the Massive Cluster Survey \citep[MACS;][]{Ebeling2001}, and has a total mass of $(1.4\pm 0.3) \times 10^{15}$~M$_{\odot}$ ($1\sigma$) within $R_{\rm 500}$ \citep{Umetsu2014}. A strong-lensing analysis of the cluster, based on Hubble Space Telescope (HST) observations, has revealed a massive merging cluster with a relatively shallow mass distribution leading to a unique lensing strength \citep{Zitrin2009}. A parametric lensing analysis \citep{Smith2009} showed that at least four merging mass halos were needed to describe the strong-lensing observables in this cluster \citep[see also][]{Rau2014,Oguri2015}, making MACS~J1149.6+2223 one of the most complex merging clusters. MACS~J1149.6+2223 is now acknowledged as one of the most powerful cosmic lenses, and is one of the clusters included  in the Cluster Lensing and Supernova Survey \citep[CLASH;][]{Postman2012}, and in the HST Frontier Fields \citep{Lotz2014}.

\begin{figure}
    \includegraphics[width=\columnwidth]{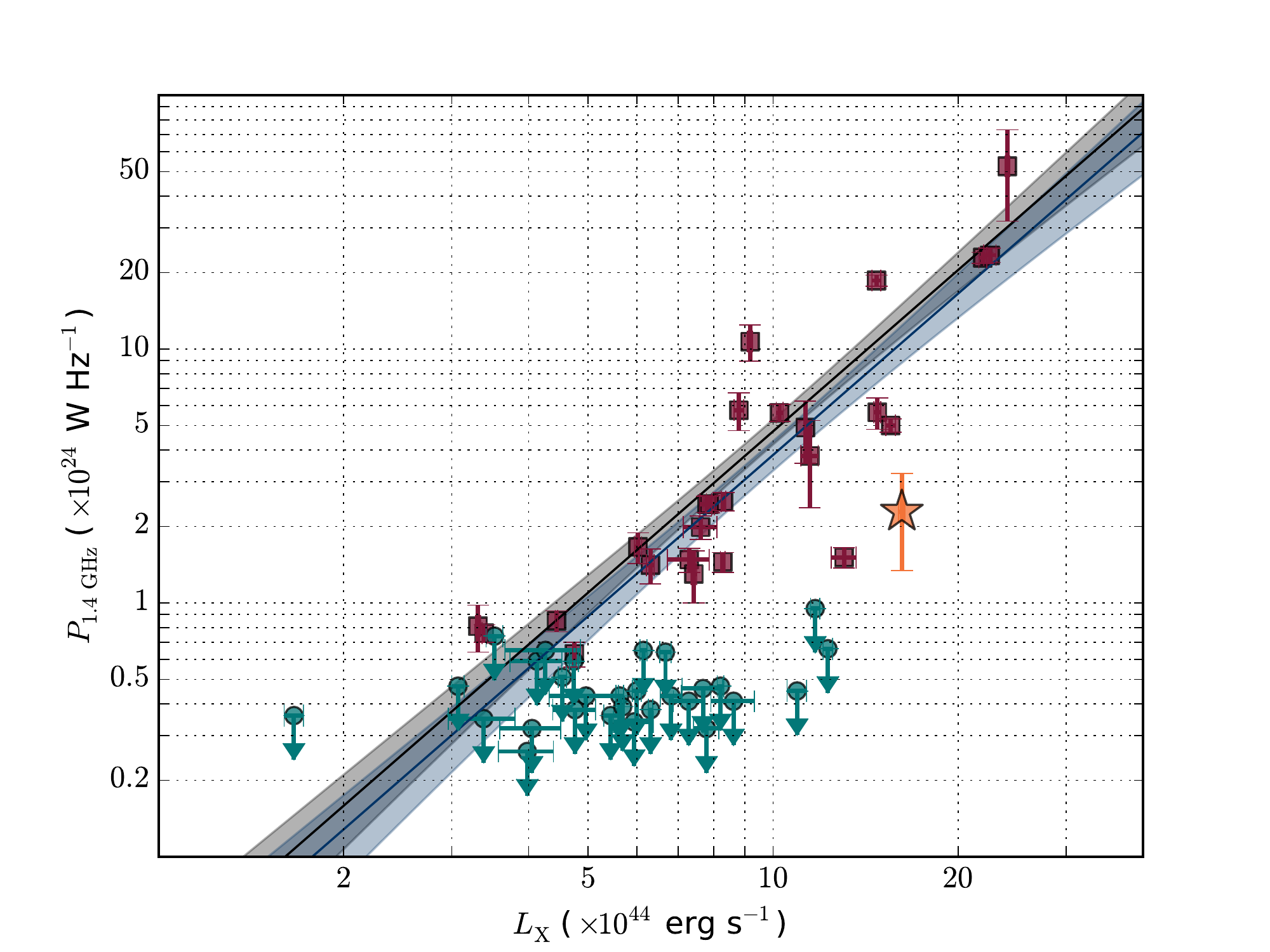}
    \caption{$L_{\rm X}--P$ diagram. Clusters with only upper limits on the radio halo power are shown in teal-colored circles, while clusters with detected halos are shown in dark red squares. MACS~J1149.6+2223 is represented with an orange star. Error bars are drawn for all points, but some are small and barely visible. Values for all clusters, with the exception of MACS~J0416.1-2403 are taken from \citet{Cassano2013}. An interactive version of the diagram is available at \url{https://goo.gl/YUAOLY} and in the online journal. \label{fig:lx-p}}
\end{figure}

Diffuse radio sources such as relics and halos are found in some merging galaxy clusters \citep[see, e.g.,][for a review]{Feretti2012}. MACS~J1149.6+2223 possibly hosts three such sources: a giant radio halo, a south-eastern (SE) relic, and possibly another relic in the west (W) \citep{Bonafede2012}. The radio halo was barely detected in archival $1.4$~GHz Very Large Array (VLA) data. The combination of the archival VLA data with 323~MHz Giant Metrewave Radio Telescope (GMRT) observations suggested that the radio halo in MACS~J1149.6+2223 has an extremely steep spectral index\footnote{The radio flux density, $S$, is proportional to $\nu^\alpha$, where $\nu$ is the frequency of the observation and $\alpha$ is the spectral index.}, $\alpha\sim -2$, indicating either a less energetic merger \citep{Brunetti2008} or an old halo in which the relativistic particles have aged. Possibly because the radio spectral index is so steep, the cluster is an outlier on the $L_{\rm X}-P$ relation \citep{Cassano2013}, as shown in Figure~\ref{fig:lx-p}. 

Here, we present results from deep \chandra\ observations of MACS~J1149.6+2223. In Section~\ref{sec:reduction}, we discuss the observations used and summarize the data reduction. In Section~\ref{sec:bkgmodeling} we detail the method used to model the foreground and background components. Sections~\ref{sec:global}$-$\ref{sec:cf} present the \chandra\ results. We discuss and summarize our results in Section~\ref{sec:summary}.

Throughout the paper, we adopt a ${\rm \Lambda CDM}$ cosmology with $H_{\rm 0}=70$~km~s$^{-1}$~Mpc$^{-1}$, $\Omega_{\rm m}=0.3$, and $\Omega_{\Lambda}=0.7$. In this cosmology, $1\arcmin$ at the redshift of MACS~J1149.6+2223 corresponds to $\approx 383$~kpc. Unless specifically mentioned otherwise, errors are quoted at the $90\%$ confidence level.


\begin{table}
   \caption{\chandra\ observations of MACS~J1149.6+2223.}
   \label{tab:observations}
   \begin{center}
   \begin{threeparttable}
      \begin{tabular}{lcccc}
         \hline
            \multirow{2}{*}{ObsID\tnote{$\star$}} \rule{0pt}{5pt} & \multirow{2}{*}{CCDs on} & \multirow{2}{*}{Starting date} & \multirow{2}{*}{Exposure} & \multirow{2}{*}{Clean} \\
             \rule{0pt}{10pt}                            &  &  &        & exposure \\
             \rule{0pt}{10pt}                            &  &  & (ks) & (ks) \\[1ex]
         \hline
           3589  &  I0, I1, I2, I3, S2 & 07-02-2003 & 20.0 & 18.3 \\
           16238 &  I0, I1, I2, I3     & 09-02-2015 & 35.6 & 30.2 \\
           16239 &  I0, I1, I2, I3, S2 & 17-01-2015 & 51.4 & 48.6 \\
            16306 & I0, I1, I2, I3, S2 & 05-02-2014 & 79.7 & 71.8 \\
            16582 & I0, I1, I2, I3, S2 & 08-02-2014 & 18.8 & 17.3 \\
            17595 & I0, I1, I2, I3     & 18-02-2015 & 69.2 & 57.1 \\
            17596 & I0, I1, I2, I3     & 10-02-2015 & 72.1 & 62.3 \\
         \hline
      \end{tabular}
      \begin{tablenotes}
           \item[$\star$] ObsID~1656 was excluded from the analysis and is not included in the table.
      \end{tablenotes}
   \end{threeparttable}
   \end{center}
\end{table}

\section{Observations and Data Reduction}
\label{sec:reduction}

MACS J1149.6+2223 was observed with \chandra\ eight times, in VFAINT mode, between 2001 and 2015, for a total of 365~ks. The observations were analyzed using CIAO~v4.7, with CALDB~v4.6.5. The data reduction steps are the same as those employed in the analysis of the \chandra\ data of MACS~J0416.1-2403 \citep{Ogrean2015}. Soft proton flares were removed from the data, point sources were subtracted, and the rescaled\footnote{The instrumental background images and spectra were rescaled such that their count rates in $10-12$~keV are the same as the count rates of the observations in the same energy band.} instrumental background was subtracted from the images and spectra.

\begin{figure}
  \includegraphics[width=\columnwidth]{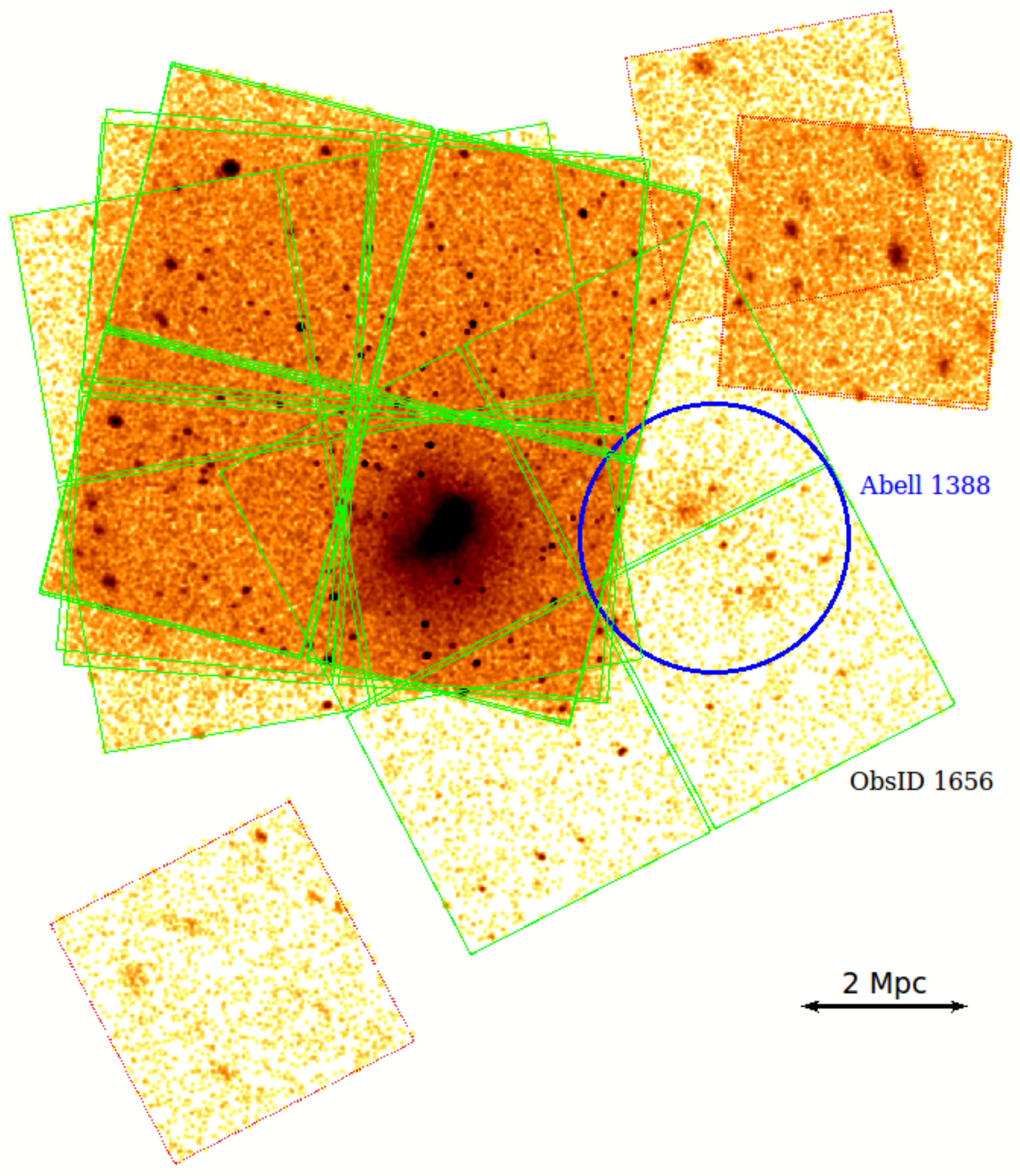}
  \caption{Footprint of the eight \chandra\ ObsIDs of MACS~J1149.6+2223. Abell~1388 \citep[$z\approx 0.18$;][]{vonderLinden2014} contaminates the FOV of ObsID~1656. A circle of radius $\approx 750$~kpc was drawn around A1388. ObsID~1656 was excluded from our analysis. \label{fig:footprint}}
\end{figure}

A footprint of the eight available ObsIDs of MACS~J1149.6+2223 is shown in Figure~\ref{fig:footprint}. In the field of view (FOV) of ObsID~1656, about $7$~arcmin west of MACS~J1149.6+2223, is the foreground galaxy cluster Abell~1388 \citep[$z\approx 0.18$;][]{vonderLinden2014}. In ObsID~1656, Abell~1388 is positioned off-axis on the ACIS-I detector, where the point spread function (PSF) is considerably degraded. Because emission from Abell~1388 covers a large part of the \chandra\ FOV in ObsID~1656, this observation is not very useful for the analysis of MACS~J1149.6+2223\footnote{We did not detect emission from Abell~1388 in the other \chandra\ ObsIDs.}. Given the relatively short exposure time of ObsID~1656 ($\sim 20$~ks), we excluded this dataset from our analysis. Therefore, the results presented here are based on 305~ks of flare-filtered \chandra\ data.

A summary of the seven ObsIDs used in this paper is presented in Table~\ref{tab:observations}.


\begin{figure*}
   \includegraphics[height=\columnwidth,angle=270]{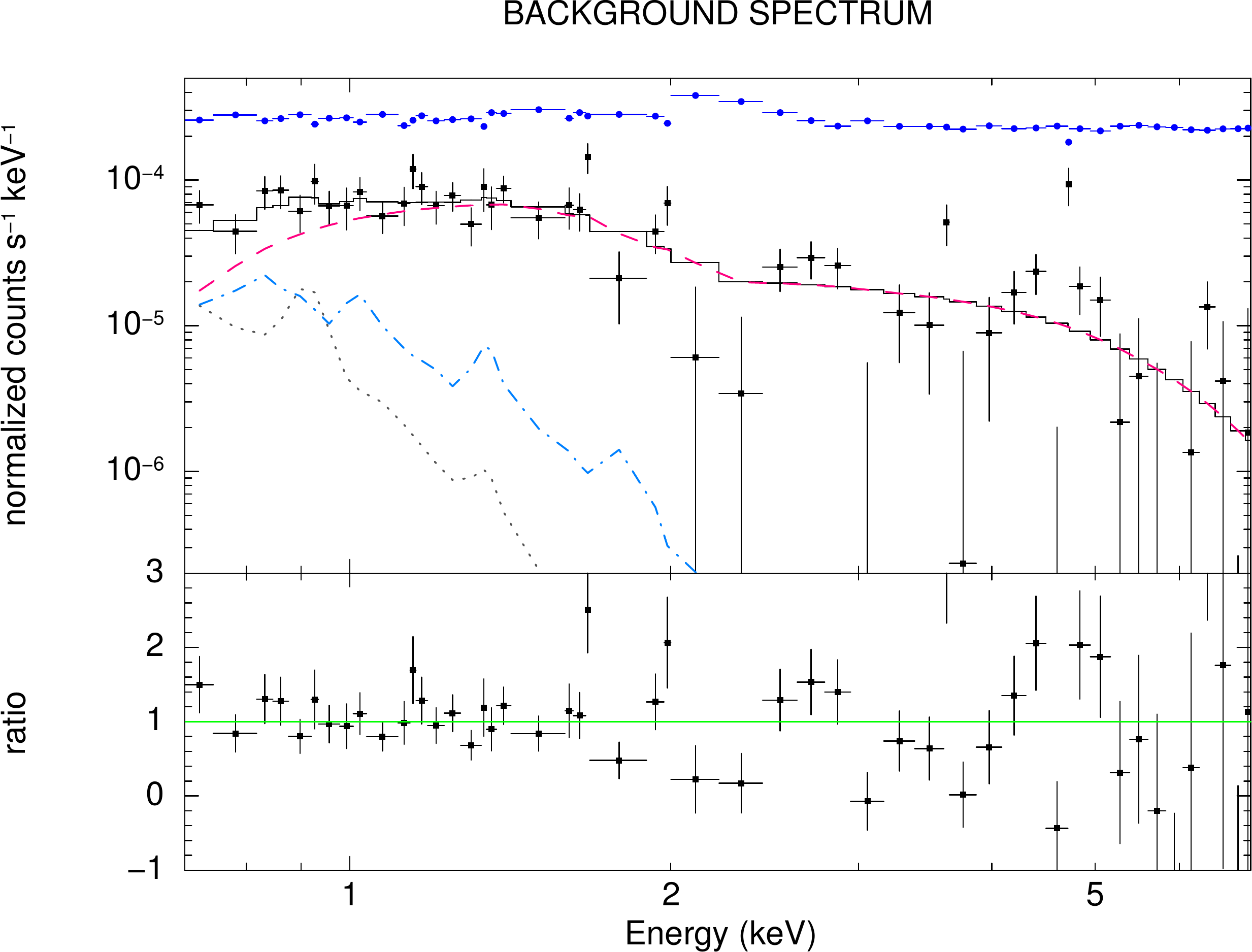}\hfill
   \includegraphics[height=\columnwidth,angle=270]{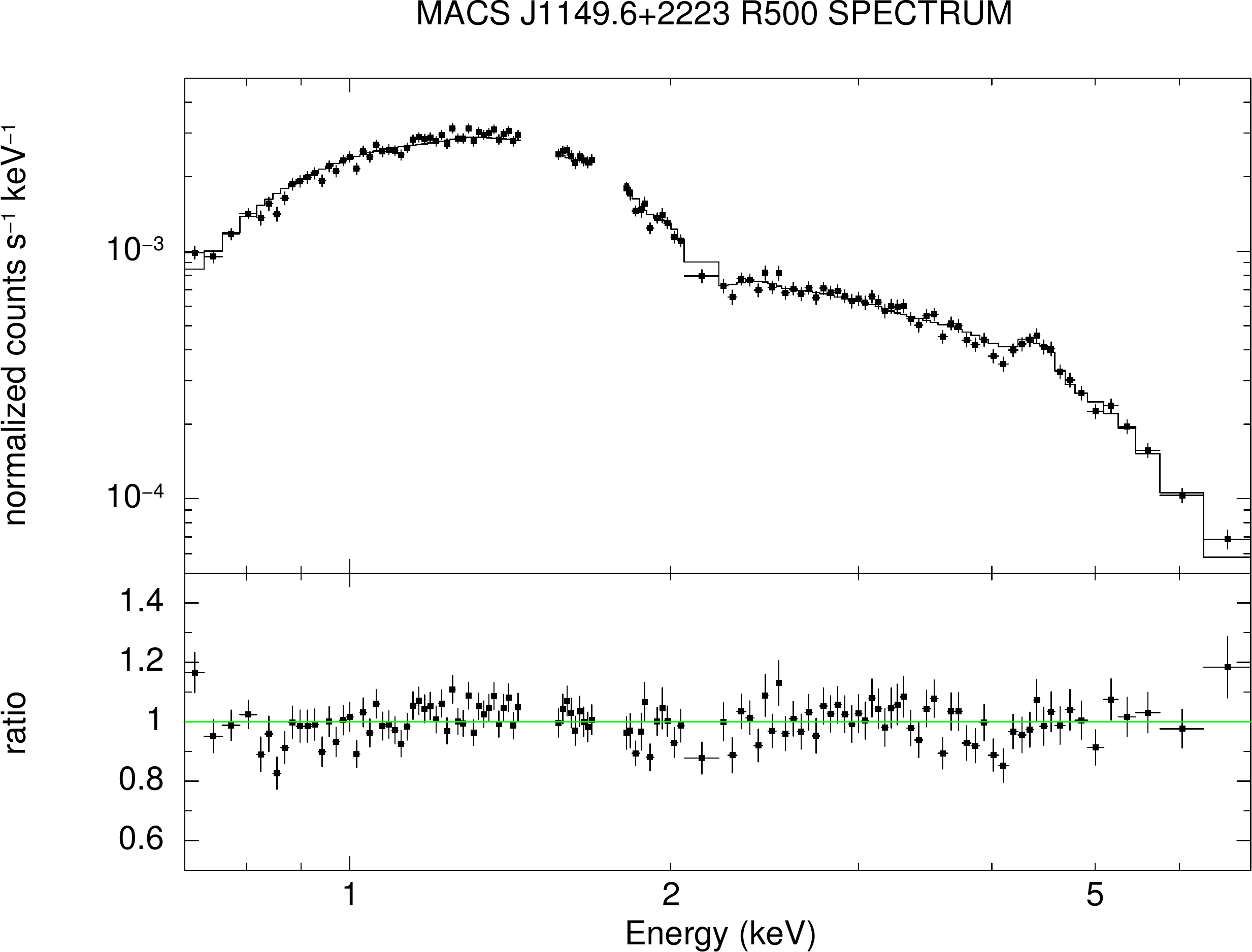}
   \caption{\emph{Left:} X-ray foreground plus background spectrum, and best-fitting model (black). The LHB, GH, and power-law components of the model are shown in gray (dotted line), light blue (dash-dotted line), and magenta (dashed line), respectively. The power-law component is an exposure-weighted average of the individual power-law components. The instrumental background spectrum is shown in dark blue. \emph{Right:} \chandra\ spectrum of the $R_{\rm 500}$ region of MACS~J1149.6+2223. For clarity, the two spectra shown here are the combined spectra from all the ObsIDs. However, the actual fits were done by fitting the individual spectra in parallel. Small energy bands ($\Delta E = 0.1$~keV) were excluded around the fluorescent instrumental lines Al~K$_\alpha$ (1.49~keV), Si~K$_\alpha$ (1.74~keV), and Au M$_{\alpha,\,\beta}$ (2.1~keV). The bottom panels show the ratio of the data to the models. \label{fig:spectra} \vspace{0.3cm}}
\end{figure*}

\section{Background Modeling}
\label{sec:bkgmodeling}

\begin{table*}
	\caption{Foreground and background spectral parameters}
	\label{tab:bkgmodel}
	\begin{center}
	     \begin{threeparttable}
		\begin{tabular}{lccc}
		     \hline
		         Spectral Component & Temperature & Power-Law Index & $0.5-2$~keV Flux \\
			                                         & $T$ & $\Gamma$ & $\mathcal{F}_{\rm 0.5-2\; keV}$ \\ 
                                                                     & (keV) &     & (erg~cm$^{-2}$~s$^{-1}$~arcmin$^{-2}$) \\[1ex]
		     \hline
		      	LHB & $0.14\pm 0.01$\tnote{$\ddagger$} & -- & $9.40_{-0.23}^{+0.52}\times 10^{-16}$\tnote{$\ddagger$} \\
			GH & $0.45_{-0.25}^{+0.31}$ & --  & $1.56_{-1.04}^{+4.60} \times 10^{-16}$ \\
			PL~ObsID~16238 & \multirow{7}{*}{--} & \multirow{7}{*}{1.41\tnote{$\dagger$}} & $8.09_{-2.18}^{+2.23} \times 10^{-16}$ \\
			PL~ObsID~16239 & & & $6.32_{-1.83}^{+1.84} \times 10^{-16}$ \\
			PL~ObsID~16306 & & & $7.44_{-1.33}^{+1.32} \times 10^{-16}$ \\
			PL~ObsID~16582 & & & $7.44_{-1.33}^{+1.32} \times 10^{-16}$ \\
			PL~ObsID~17595 & & & $5.25_{-1.19}^{+1.18} \times 10^{-16}$ \\
			PL~ObsID~17596 & & & $5.25_{-1.19}^{+1.18} \times 10^{-16}$ \\
			PL~ObsID~3589  & & & $9.56_{-2.03}^{+2.38} \times 10^{-16}$ \\
	     	     \hline
		\end{tabular}
		\begin{tablenotes}
			\item[$\dagger$] Fixed parameter.
			\item[$\ddagger$] \rosat\ best-fitting value. Fixed \chandra\ spectral parameter.
		\end{tablenotes}
	     \end{threeparttable}
	\end{center}
\end{table*}

The analysis of features in the outskirts of galaxy clusters, where the ICM surface brightness is low, necessitates a careful treatment of the background components. We modeled the X-ray foreground and background using spectra extracted from regions free of ICM emission. The stowed background was subtracted from the spectra, and the remaining emission was described as the sum of unabsorbed Local Hot Bubble (LHB) emission, absorbed Galactic Halo (GH) emission, and absorbed emission from unresolved X-ray point sources. The spectral parameters of the Local Hot Bubble in the direction of MACS~J1149.6+2223 were determined from a \rosat\ All-Sky Survey (RASS) background spectrum extracted in an annulus with radii $0.3-1.3$~degrees around the cluster center ($0.3$~degrees corresponds approximately to $3R_{\rm 200}$, and the inner radius was chosen to be so large in order to avoid contamination from Abell~1388 in the RASS spectrum). The temperature and normalization of the Local Hot Bubble were kept fixed to the best-fitting RASS values when fitting the \chandra\ spectra. The Galactic Halo parameters were linked for the seven \chandra\ spectra. ObsIDs 17595 and 17596, and ObsIDs 16306 and 16582 cover essentially the same region of the sky. For these pairs of observations, we kept the normalizations of the power-laws (PL) describing unresolved point sources free but linked. The power-law normalizations were free and unlinked for the other ObsIDs. The indices of the power-laws were fixed to $1.41$ \citep{DeLucaMolendi2004}.

The spectra were binned to have a minimum of 1 count per bin, and fitted using the extended C-statistic \citep{Cash1979,Wachter1979} in \textsc{XSpec}~v12.8.2. We restricted the fit to the energy band $0.7-7$~keV. We assumed the solar abundance table of \citet{feld92}, and \citet{vern96} photoelectric absorption cross-sections. The hydrogen column density in the direction of the cluster was fixed to a value of $2.06\times 10^{20}$~cm$^{-2}$, which represents the sum of atomic and molecular hydrogen column densities\footnote{\url{http://www.swift.ac.uk/analysis/nhtot/index.php}} in a circle with a radius of $1$~degree around the cluster center \citep{LAB,SWIFT}. In Table~\ref{tab:bkgmodel}, we summarize the best-fitting sky foreground and background model. The model is shown in Figure~\ref{fig:spectra}.


\section{Global X-ray Properties}
\label{sec:global}

\begin{figure*}[t]
    \includegraphics[width=\textwidth]{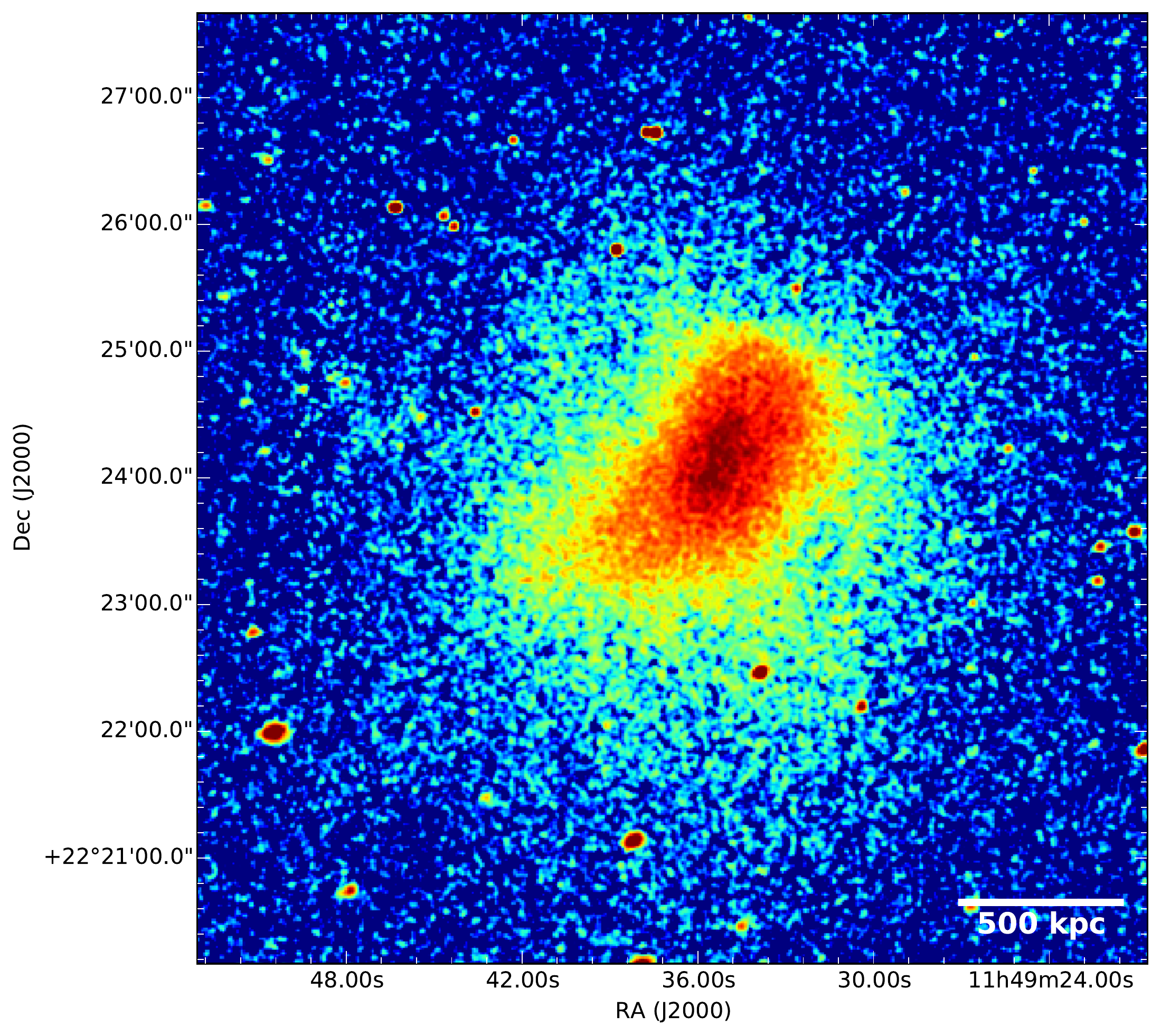}
    \caption{\chandra\ surface brightness map in the energy band $0.5-3$~keV. The image was exposure-corrected, vignetting-corrected, instrumental background-subtracted, and smoothed with a Gaussian kernel of width $1$~arcsec. \label{fig:sxmap} \vspace{0.3cm}}
\end{figure*}

The surface brightness map of the cluster is shown in Figure~\ref{fig:sxmap}. The cluster is elongated in the NW-SE direction, and has a single surface brightness concentration. { There is little substructure visible in the ICM, unlike in some of the other Frontier Field clusters such as MACS~J0717.5+3745 and MACS~J0416.1-2403 \citep[e.g.,][]{vanWeeren2015,Ogrean2015}}. The X-ray emission is asymmetrical: more elongated towards SE than towards NW, and brighter and more extended in the SW than in the NE. A hint of a surface brightness edge can be seen N-NE of the cluster center; the edge is discussed in Sections~\ref{sec:edges} and\ref{sec:cf}. 

To measure the global properties of the cluster, we extracted spectra in a circle with radius $R_{\rm 500}= 1.6$~Mpc \citep{Sayers2013} around the cluster center, { chosen to be at ${\rm RA} = 11^{\rm h}\, 49^{\rm m}\, 36^{s}$ and ${\rm Dec} = +22^{\circ}\, 24^{\prime}\, 05^{\prime\prime}$.} Stowed background spectra extracted from the same regions were subtracted from the total spectra. The remaining emission was modeled as the sum of sky background and ICM signal. The sky background parameters were fixed to the values in Table~\ref{tab:bkgmodel}. The ICM was modeled with a single-temperature thermal component \citep[APEC;][]{Smith2001,Foster2012}. The cluster redshift was fixed to $0.544$, while the other parameters of the APEC model were free in the fit. We determined that MACS~J1149.6+2223 has an average temperature $T_{\rm 500} = 10.73_{-0.43}^{+0.62}$~keV, a metallicity $Z_{\rm 500} = 0.20\pm 0.05$~$Z_{\sun}$, and a rest-frame luminosity $L_{\rm 500,\; [0.1-2.4\,\,keV]} = (1.62\pm 0.02)\times 10^{45}$~erg~s$^{-1}$. The $R_{\rm 500}$ spectrum and the best-fitting model are shown in Figure~\ref{fig:spectra}.


\section{ICM Temperature Distribution}
\label{sec:tmap}

\begin{figure}
    \includegraphics[width=\columnwidth]{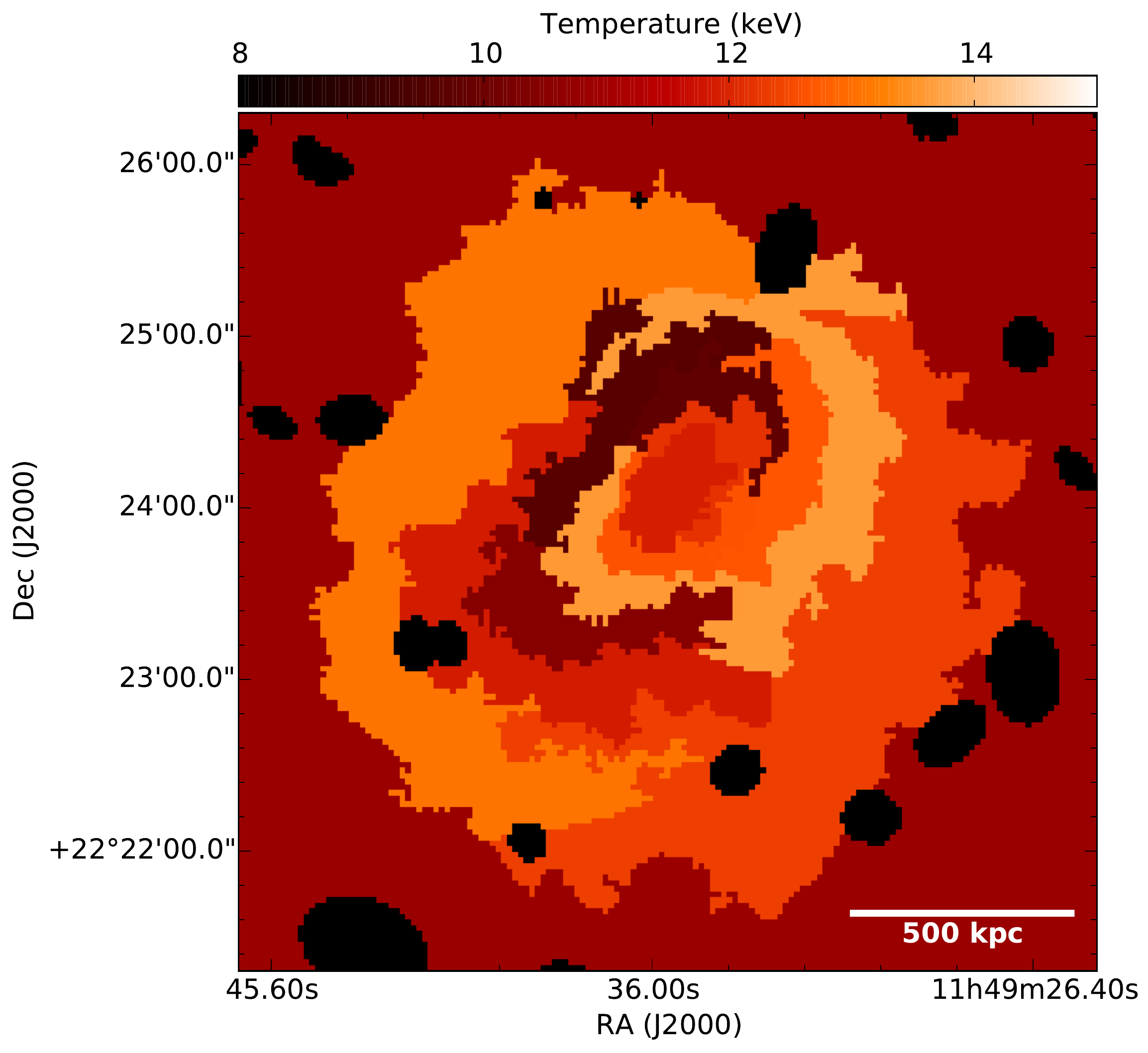}
    \caption{Temperature map of MACS~J1149.6+2223. Black ellipses mark regions from which point sources have been removed. The outermost spectral region is not fully shown in this figure; its actual area is about twice as large as shown. \label{fig:tmap}\vspace{0.3cm}}
\end{figure}

The \textsc{contbin} code \citep{Sanders2006} was used to divide the surface brightness map of MACS~J1149.6+2223 in regions with 3600~counts from the ICM and sky background combined. The regions follow the contours on a surface brightness map adaptively smoothed to achieve ${\rm SNR} \approx 10$. The temperature distribution was mapped by extracting spectra from each of the regions, and fitting them with a single-temperature absorbed APEC component. The instrumental background was subtracted from the data, while the sky background was kept fixed to the parameters listed in Table~\ref{tab:bkgmodel}. The cluster redshift was fixed to $0.544$, while the metallicity was fixed to the $R_{\rm 500}$ value of $0.20$~Z$_\sun$. The resulting temperature map is shown in Figure~\ref{fig:tmap}. An interactive version, which includes the statistical errors on the measurements at a confidence level of $90\%$, is available at \url{https://goo.gl/1CftGu} and in the online journal.

Throughout the ICM, the temperature is consistent with $\sim 10$~keV. Temperature jumps between adjacent regions in Figure~\ref{fig:tmap} would suggest the presence of shocks or cold fronts in the ICM. However, we do not detect statistically significant (at $90\%$ confidence level) temperature jumps. At $1\sigma$ significance, there is a hint of a cold front $\sim 0.7-1$~arcmin N-NE of the cluster center. This feature will be examined more carefully in the next section.


\section{Surface Brightness Edges}
\label{sec:edges}

\subsection{Edges near the Radio Relics}

Surface brightness edges indicate underlying density discontinuities in the ICM that correspond either to shock fronts or to cold fronts, depending on the direction of the temperature jump \citep[e.g.,][]{Markevitch2007}. MACS~J1149.6+2223 hosts one radio relic in the SE, and a radio relic candidate in the NW \citep{Bonafede2012}. Because radio relics are believed to result from particle acceleration at shock fronts \citep[e.g.,][]{Markevitch2002,Markevitch2006,Russell2012}, we searched for surface brightness edges near the locations of the SE relic and the NW candidate relic. Unfortunately, in all the \chandra\ ObsIDs, the cluster was positioned near the edge of the FOV, in the SW corner of the detector. This positioning means that the two radio sources are at the very edge of the FOV, and all the X-ray point sources near the diffuse radio sources are significantly extended. Because of this, even if surface brightness edges were present near the diffuse radio sources, there are not enough counts available in the putative pre-shock regions to allow us to detect these edges.

\subsection{A N-NE Surface Brightness Edge}

Sharp surface brightness edges are not immediately visible in Figure~\ref{fig:sxmap}. To search for weak edges, we examined the unsharp masked image of the cluster. The unsharp masked image, shown in Figure~\ref{fig:ds}, was created by smoothing the surface brightness map with two Gaussians of widths $3$ and $15$~arcsec, and dividing their difference to the map with the larger smoothing. In the unsharp masked image, a strong gradient is seen N-NE of the cluster center.\footnote{We tried various smoothing scales between $2$ and $25$~arcsec, but no additional substructure was revealed.} This gradient could indicate a surface brightness edge.

\begin{figure}
   \centering
    \includegraphics[width=0.91\columnwidth]{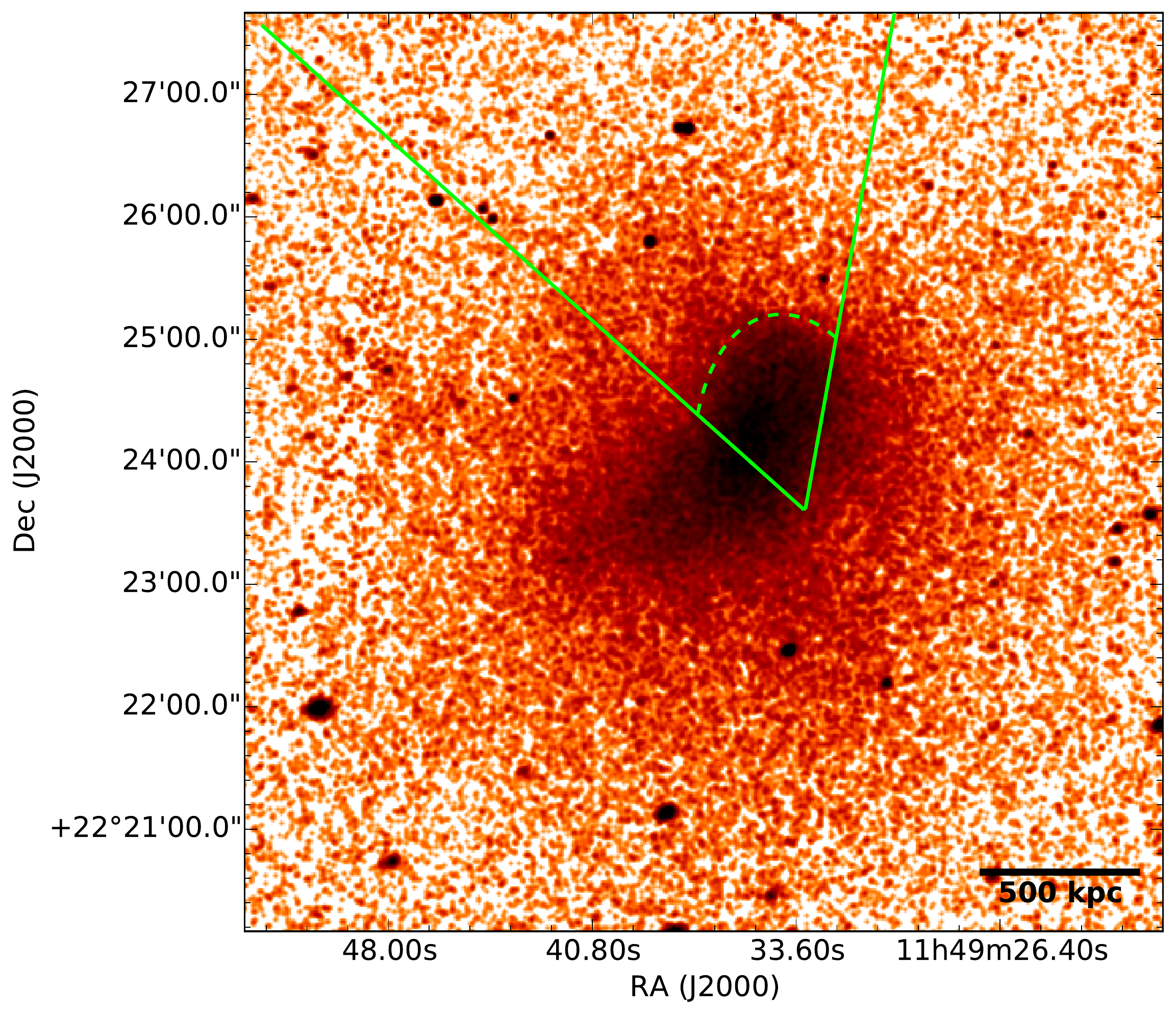}
    \includegraphics[height=\columnwidth,angle=270]{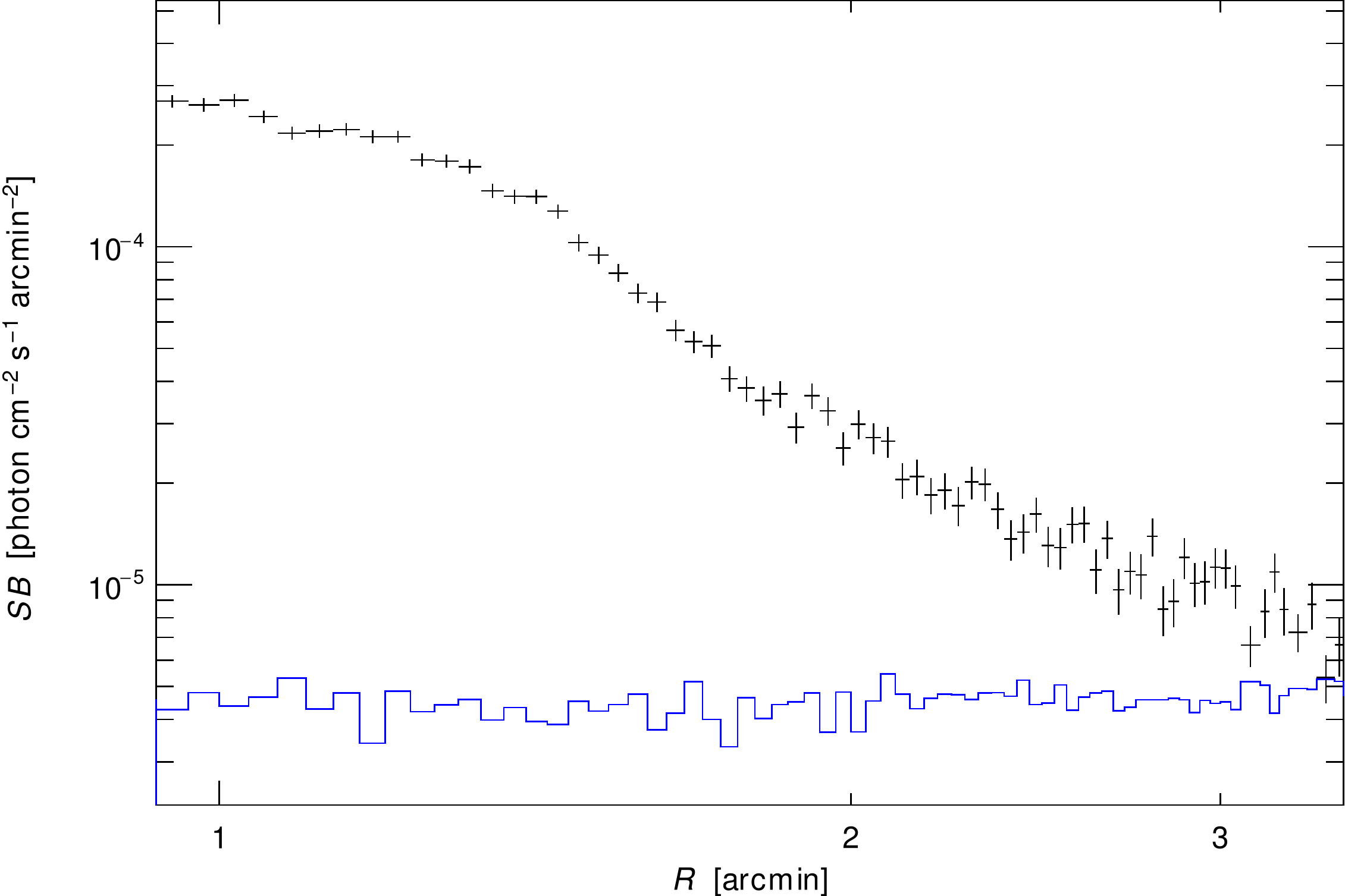}\\
    \includegraphics[height=\columnwidth,angle=270]{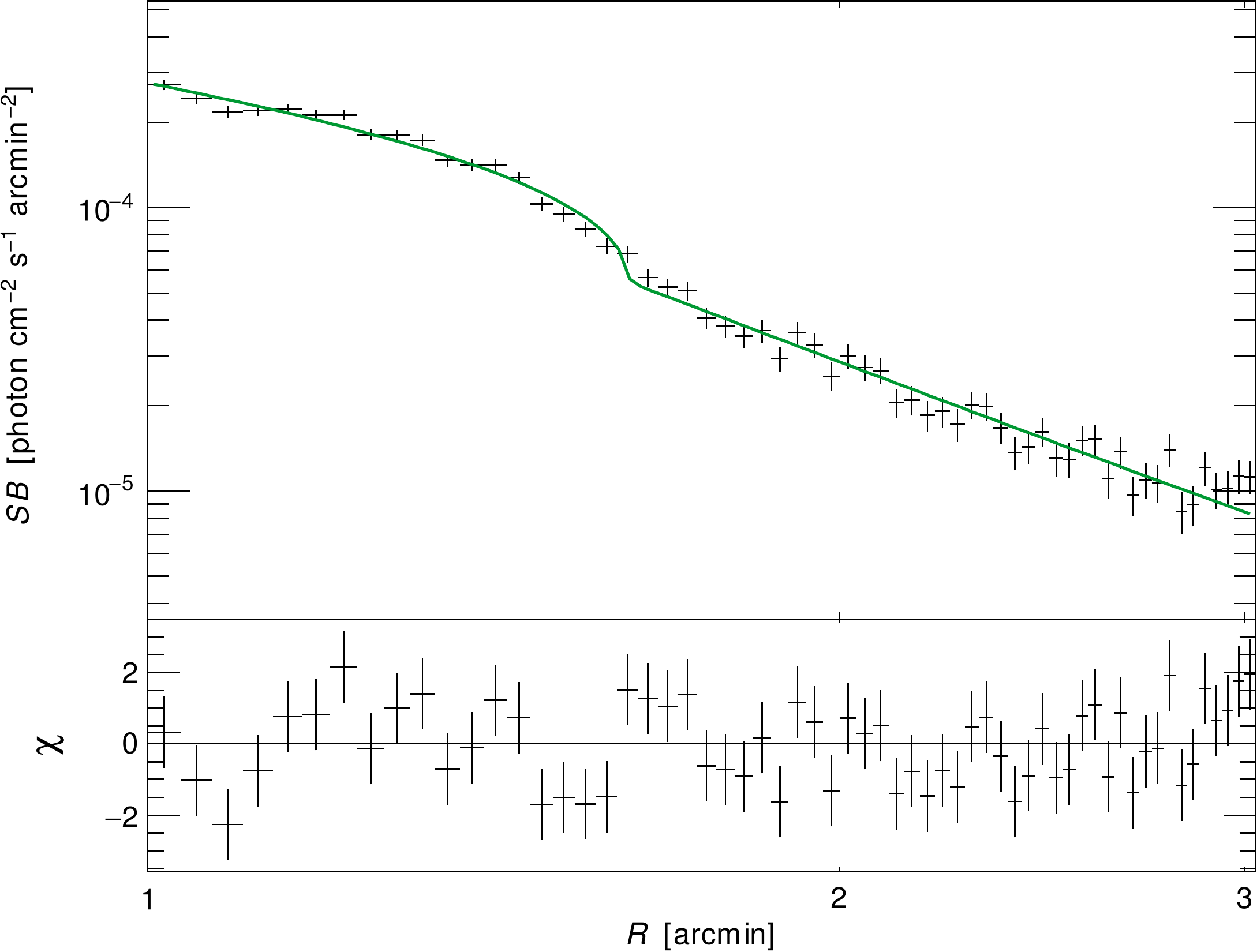}
    \caption{\emph{Top:} Sector used to model the surface brightness profile across the N-NE brightness edge. The dashed arc shows the location of the surface brightness edge. \emph{Middle:} $0.5-4$~keV surface brightness profile in the sector shown above, binned to a uniform SNR of 5. This high binning is done here only for clarity; a more finely binned profile was used for determining the best-fitting parameters (see text for details). The instrumental background profile is shown in blue. \emph{Bottom:} Best-fitting broken power-law density model fitted to the surface brightness edge seen in the sector above. The surface brightness profile is shown with the instrumental background subtracted. \label{fig:confirmed-jumps} \vspace{0.3cm}}
\end{figure}

We studied the putative surface brightness edge by extracting a surface brightness profile in an elliptical sector that approximately follows the X-ray surface brightness contours near the possible edge. The sector and the surface brightness profile across the putative edge revealed by the unsharp masked image are shown in Figure~\ref{fig:confirmed-jumps}. The surface brightness profile around the putative edge was fitted with three density models: a power-law density model, a kink density model, and a broken power-law density model. The density models were squared and integrated along the line of sight assuming isothermal plasma and a prolate spheroidal geometry. The broken power-law density model was parametrized as:
\begin{eqnarray}
        n(r) = \begin{cases} C\,n_{\rm {0}} \left(\frac{r}{r_{\rm d}}\right)^{-\alpha}\,, & \mbox{if } r \le r_{\rm d} \\ n_{\rm {0}} \left(\frac{r}{r_{\rm d}}\right)^{-\beta}\,, & \mbox{if } r > r_{\rm d} \end{cases} \,,
\end{eqnarray}
where $n$ is the electron number density, $C$ is the density compression, and $r_{\rm d}$ is the radius of the density jump. The kink model is a broken power-law density model with $C=1$. The sky background level was calculated by fitting a constant to the surface brightness at large radii, where no ICM emission is seen. The sky background was consequently fixed when fitting the ICM surface brightness. The systematic uncertainties on the ICM density parameters were calculated by varying the sky background within the $90\%$ confidence interval, and added in quadrature to the $90\%$-level statistical uncertainties. We assumed that the sky background does not vary spatially outside the $90\%$ confidence interval.

We binned the profile to have approximately $20$ counts/bin, and fitted it with a modified version of the \textsc{proffit} package \citep{Eckert2012}. We used Cash statistics, and compared the various fits using the likelihood-ratio test.

In the profile along the N-NE direction (Figure~\ref{fig:confirmed-jumps}), a weak edge is visible at a radius of $\sim 1.5$~arcmin. The sky background in this sector was modeled by fitting a constant to the radius range $6-12$~arcmin, and found to be $5.16_{-0.74}^{+0.75}\times 10^{-7}$~photons~cm$^{-2}$~s$^{-1}$~arcmin$^{-2}$. A broken power-law density model fit to radius range $1-3$~arcmin yielded a best-fitting density compression $C=1.57_{-0.12}^{+0.09}$ (statistical plus systematic uncertainty, calculated as described above). The broken power-law model is shown in the bottom panel of Figure~\ref{fig:confirmed-jumps}. The best-fitting parameters are summarized in Table~\ref{tab:jumps}.\footnote{The best-fitting density compression does not change significantly if we assume that the cluster is an oblate spheroid.}

\begin{table*}
   \caption{Best-fitting density model parameters for the profiles shown in Figures~\ref{fig:confirmed-jumps} and \ref{fig:hint-jump}.}
   \label{tab:jumps}
   \begin{center}
   \begin{threeparttable}
      \begin{tabular}{lccccccc}
         \hline
            \rule[2ex]{0pt}{2ex}\textsc{{N$-$NE sector (inner edge)}} \\[2ex]
         \hline
            \multirow{2}{*}{Model} \rule{0pt}{10pt} & \multirow{2}{*}{$\alpha$} & \multirow{2}{*}{$\beta$} & $r_{\rm d}^{\star}$ & \multirow{2}{*}{$C$} & $SB_{\rm skybkg}$ & \multirow{2}{*}{${\rm \Delta C}$-stat} & \multirow{2}{*}{${\rm CL}^\ddagger$} \\
              \rule{0pt}{8pt}     &         &        & (kpc) &        & (photons~cm$^{-2}$~s$^{-1}$~arcmin$^{-2}$) &   &    \\[1ex]
         \hline
            Power-law & $3.50\pm 0.06$ & -- & -- & -- & \multirow{3}{*}{$(5.16_{-0.74}^{+0.75}) \times 10^{-7}$} & 0 & -- \\
            Kink             & $0.25_{-0.64}^{+0.45}$ & $2.48\pm 0.05$ & $498\pm 15$ & -- &   & 175 & $\approx 100\%$   \\
            Broken power-law & $0.98_{-0.33}^{+0.15}$ & $2.06_{-0.09}^{+0.15}$ & $620_{-23}^{+8}$ & $1.57_{-0.12}^{+0.09}$ &   & 250 & $\approx 100\%$ \\ 
         \hline
            \rule[2ex]{0pt}{2ex}\textsc{{NE sector (outer edge)}} \\[2ex]
         \hline
            Power-law & $2.28_{-0.30}^{+0.32}$ & -- & -- & -- & \multirow{3}{*}{$(6.08_{-1.00}^{+1.02}) \times 10^{-7}$} & 0 & --   \\
            Kink             & $0.92_{-0.69}^{+0.51}$ & $2.30_{-0.52}^{+0.77}$ & $563_{-107}^{+161}$ & -- &   & 6.6 & $96.3\%$  \\
            Broken power-law & $0.54_{-0.70}^{+0.61}$ & $1.51_{-0.81}^{+0.93}$ & $605_{-27}^{+31}$ & $>1.13$ &   & 10.2 & $98.3\%$  \\ 
         \hline
      \end{tabular}
      \begin{tablenotes}
	\item[$\star$] Radii are measured from the center of the sectors, not from the cluster centers.
	\item[$\ddagger$] Confidence level at which a model describes the data better than a power-law density model.
      \end{tablenotes}
   \end{threeparttable}
   \end{center}
\end{table*}

\subsection{Hints of a Second NE Edge}

\begin{figure*}
   \centering
    \includegraphics[width=0.33\textwidth]{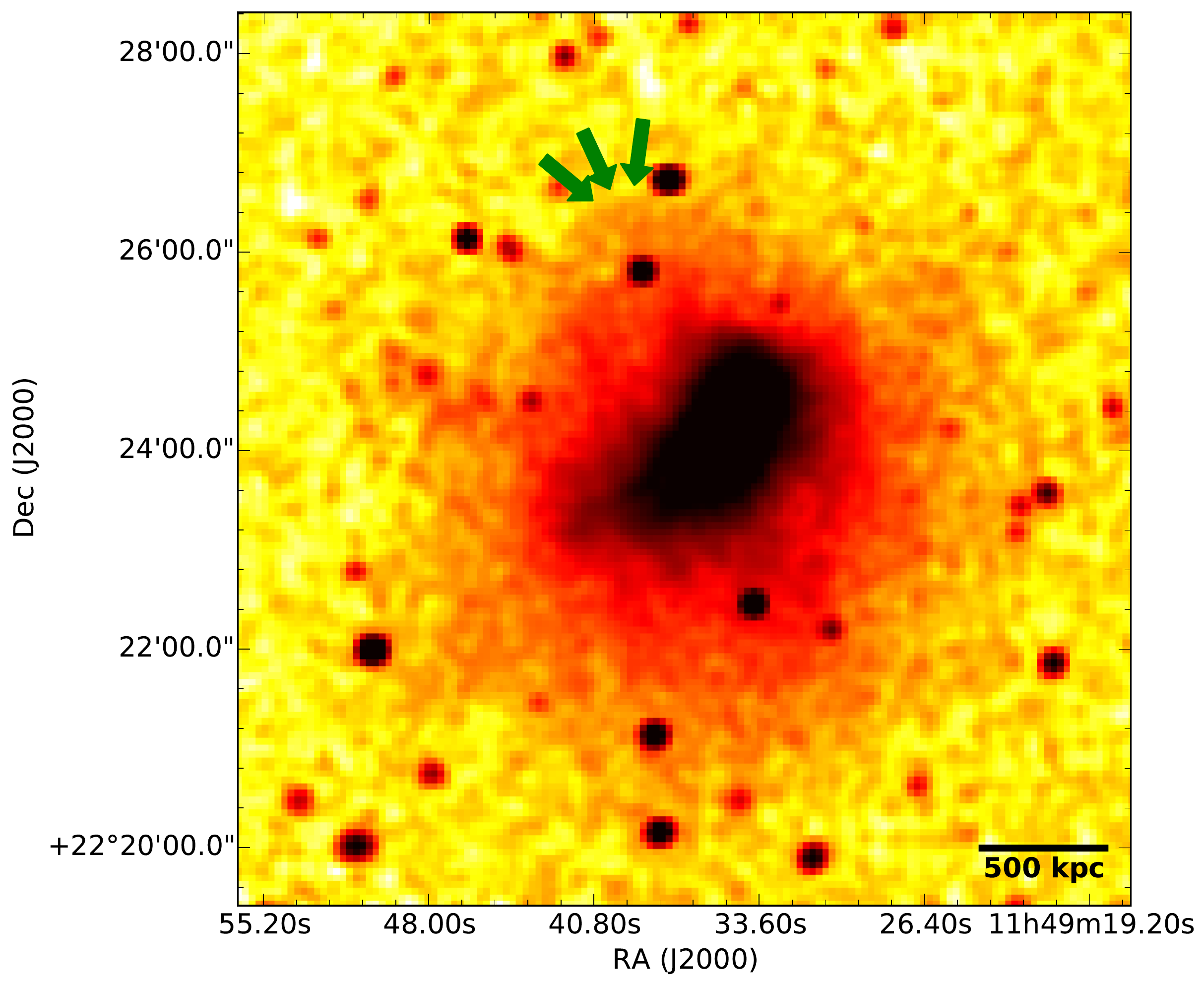}\hfill
    \includegraphics[width=0.33\textwidth]{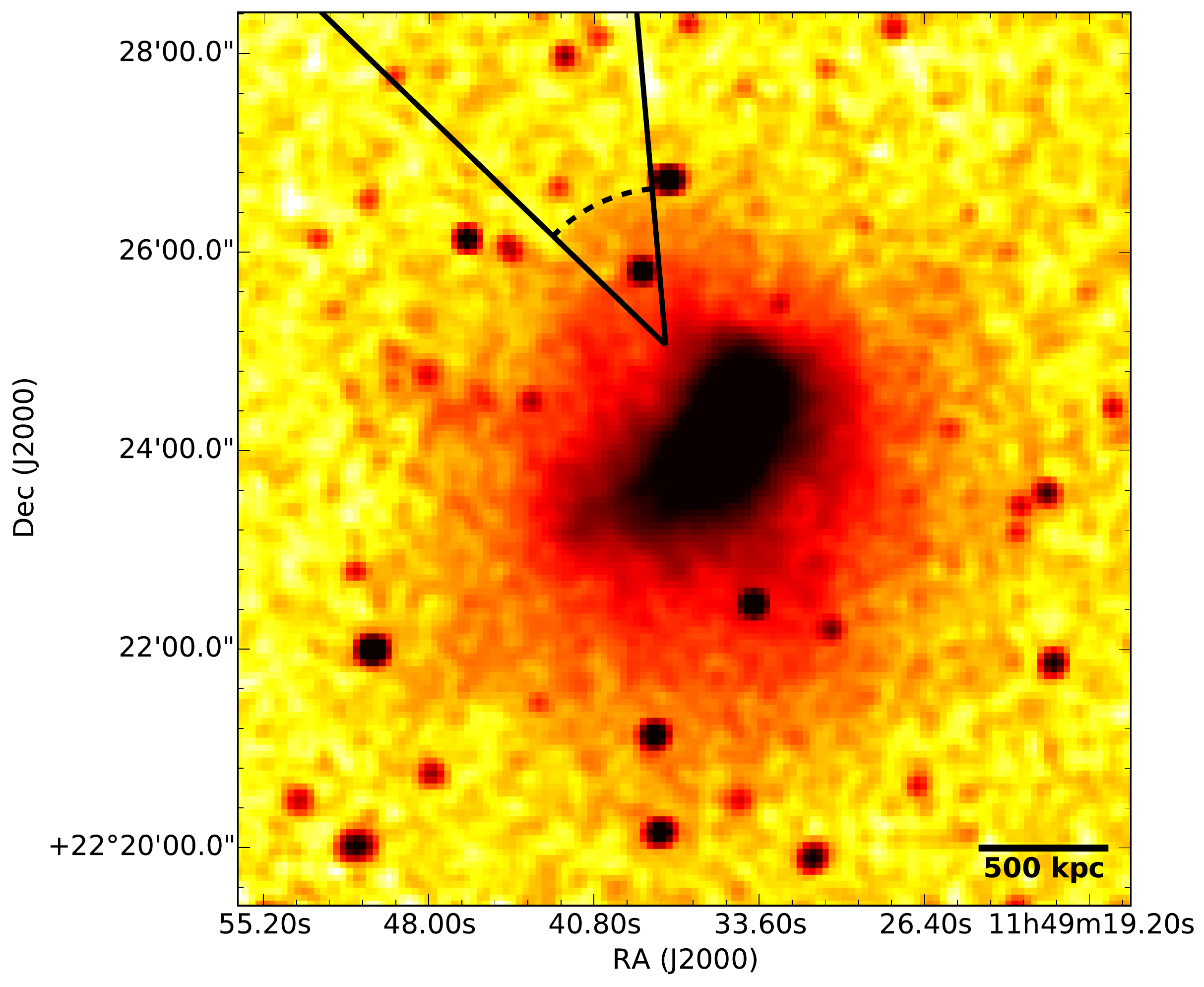}\hfill
    \includegraphics[width=0.33\textwidth]{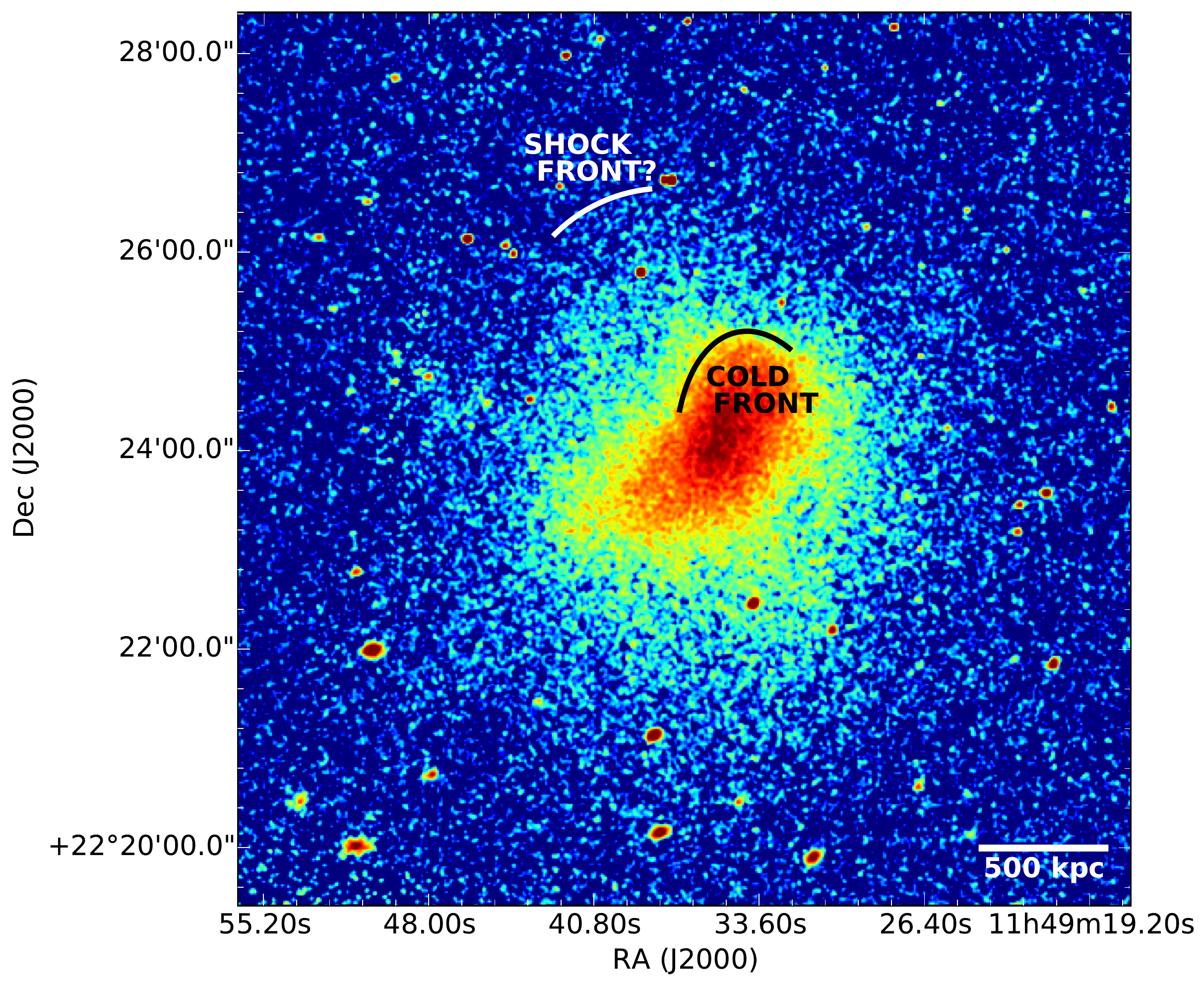}\\
    \includegraphics[height=\columnwidth,angle=270,valign=b]{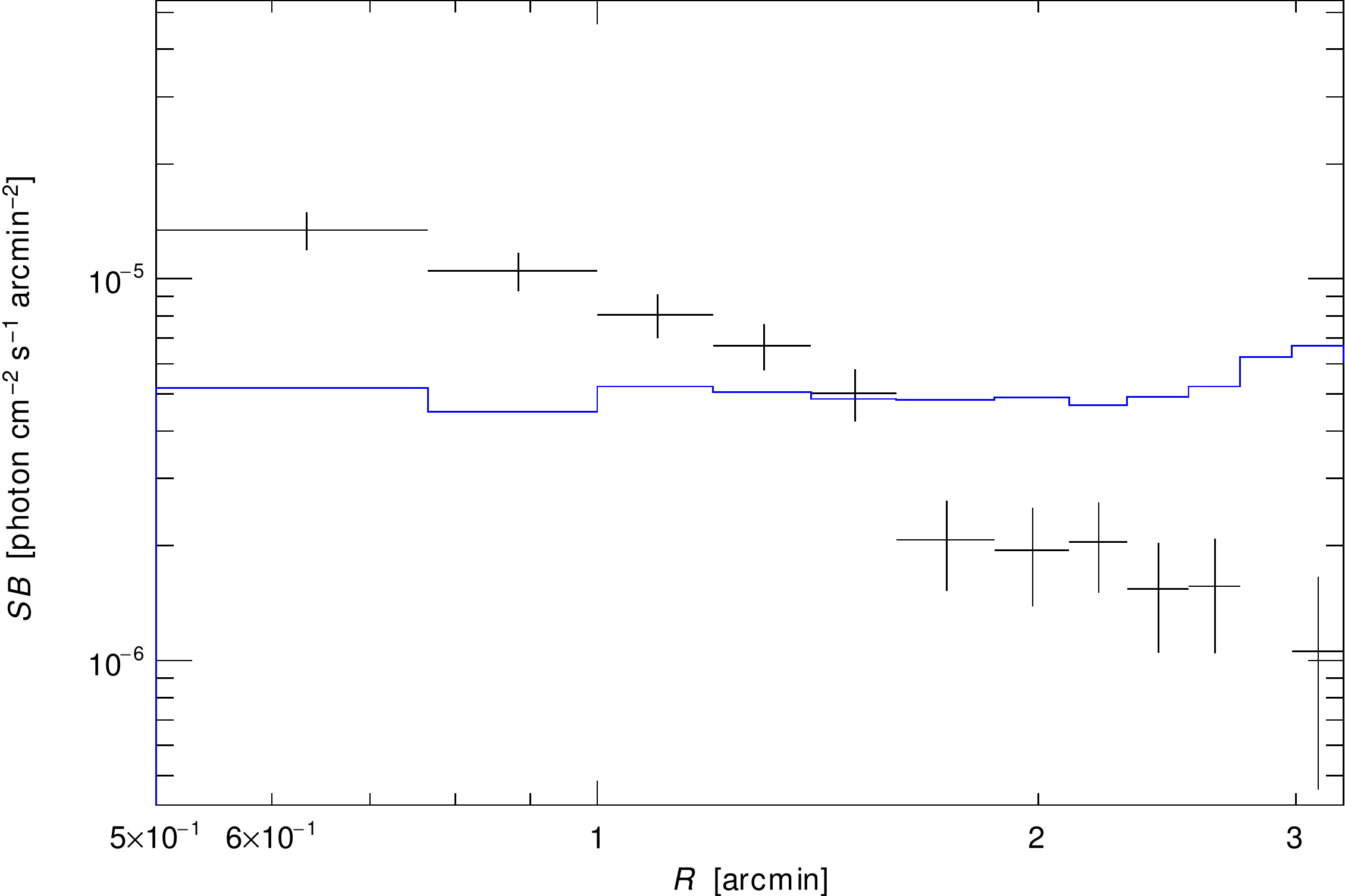}\hfill
    \includegraphics[height=\columnwidth,angle=270,valign=b]{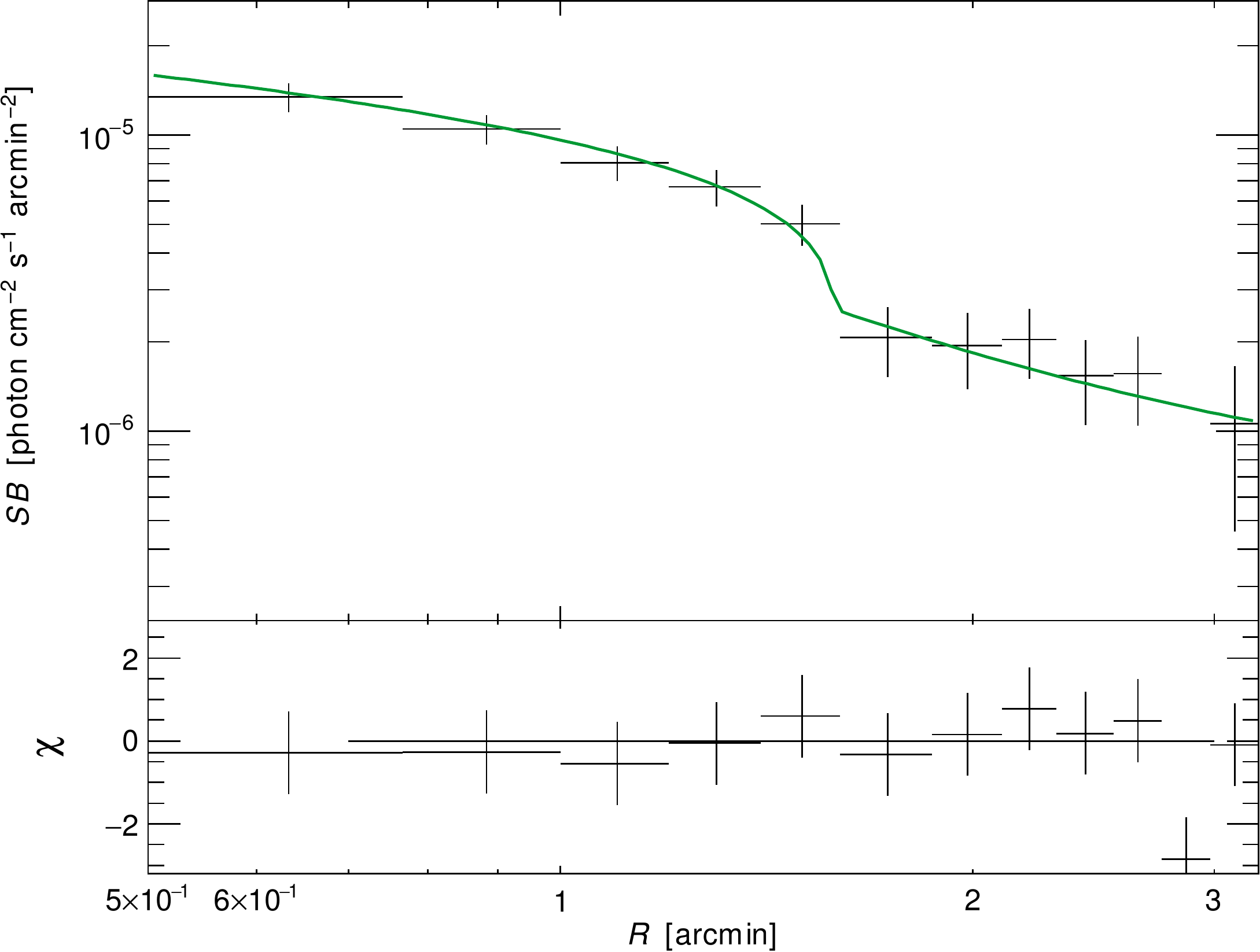}
    \caption{\emph{Top-left:} $0.5-4$~keV surface brightness map, binned by 8 and smoothed with a Gaussian of width $2$~arcsec. The map was scaled to reveal a possible surface brightness edge located $\sim 1$~Mpc from the cluster center. The position of the putative edge is indicated with arrows. \emph{Top-middle:} Sector used to model the surface brightness profile across possible NE brightness edge. Dashed line shows the location of the surface brightness edge indicated in the top-left panel. \emph{Top-right:} Same surface brightness map as in Figure~\ref{fig:sxmap}, with the positions of the cold front and putative shock front marked. \emph{Bottom-left:} $0.5-4$~keV surface brightness profile in the sector shown above, binned to 150 counts/bin. This high binning is done here only for clarity; more finely binned profiles were used for determining the best-fitting parameters (see text for details). The instrumental background profile is shown in blue. \emph{Bottom-right:} Best-fitting broken power-law density model fitted to the possible surface brightness edge seen $\sim 1$~Mpc NE of the cluster center. The surface brightness profiles are shown with the instrumental background subtracted. \label{fig:hint-jump} \vspace{0.3cm}}
\end{figure*}

In analyzing the N-NE edge, we detected a hint of an additional surface brightness edge in the NE, approximately 1~Mpc away from the cluster center. A surface brightness map scaled to reveal this putative edge is shown in the top-left panel of Figure~\ref{fig:hint-jump}. The surface brightness profile across the putative edge and the sector from which it was extracted are also shown in Figure~\ref{fig:hint-jump}. The best-fitting broken power-law model that describes the profile has a density compression $C>1.13$ at the $90\%$ confidence level (when systematic uncertainties associated with the sky background level are also considered); only a lower limit could be set on this parameter. The density compression is $>1$ at a confidence level of $94\%$.\footnote{We tried various sectors in which the putative edge was positioned between $300-1000$~kpc, but the lower limit on the density compression was always consistent with $C>1.1$ and the location of the discontinuity was essentially unchanged.} The best-fitting density parameters are summarized in Table~\ref{tab:jumps}, and the broken power-law fit is shown in the bottom-right panel of Figure~\ref{fig:hint-jump}.


\section{The Nature of the Brightness Edges}
\label{sec:cf}

The count statistics beyond the putative outer NE edge are too poor (signal-to-noise ratio [SNR] $\sim 5$) to allow a temperature measurement, so we were not able to determine the nature of the outer NE edge. It could be either a shock front or a cold front.

\begin{figure}
   \includegraphics[width=\columnwidth]{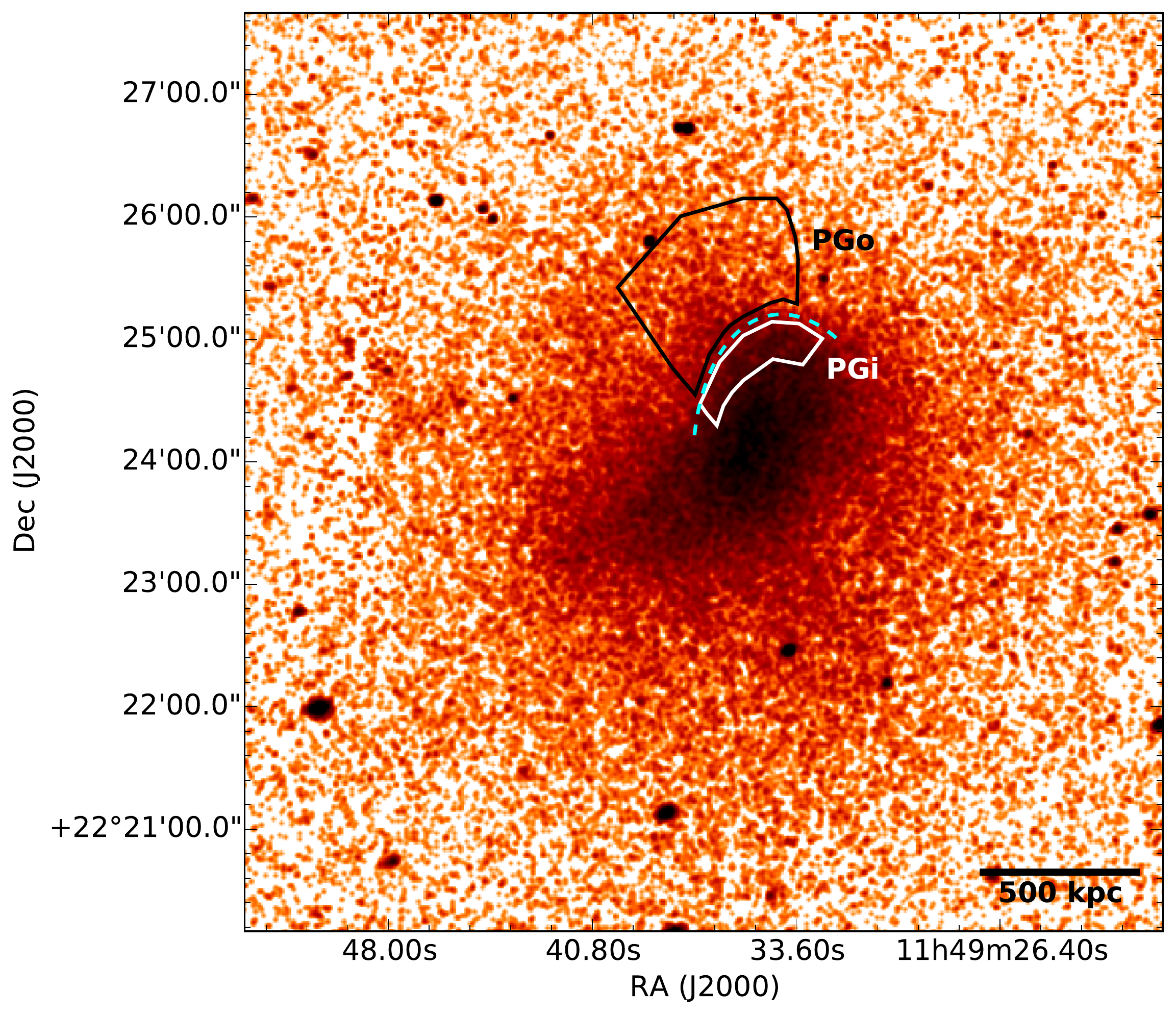}
   \caption{Regions used to measure the temperatures on both sides of the N-NE brightness edge. The dashed cyan arc marks the location of the density discontinuity. \label{fig:cf}}
\end{figure}

To determine the nature of the inner N-NE density discontinuity, we extracted spectra from two polygons (PGo -- outer region; and PGi -- inner region) on either side of the discontinuity. The polygons are shown in Figure~\ref{fig:cf}. PGo has approximately 2200 ICM counts (signal-to-noise ratio $\sim 40$), while PGi has approximately 2900 ICM counts (signal-to-noise ratio $\sim 50$). The best-fitting temperatures are $T_{\rm PGi} = 8.23_{-1.01}^{+1.46}$~keV and $T_{\rm PGo} = 9.88_{-1.94}^{+3.13}$~keV, so the temperatures in PGo and PGi are consistent with each other. 

If the inner surface brightness edge is a shock front travelling outwards through the ICM, then the Mach number calculated from the density compression using the Rankine-Hugoniot jump conditions is $\mathcal{M}=1.39_{-0.09}^{+0.07}$. This Mach number corresponds to a temperature jump $T_{\rm PGi}/T_{\rm PGo} = 1.38_{-0.09}^{+0.07}$. For a post-shock temperature $T_{\rm PGi} = 9.7$~keV (statistically, the worst case scenario; this is the upper $90\%$ confidence limit on the temperature in PGi), this temperature jump would imply $T_{\rm PGo} \approx 7.0\pm 0.4$~keV. We refitted the PGo spectrum with the ICM temperature fixed first to 7.4~keV (again, statistically, the worst case scenario) and then fixed to 9.9~keV. The difference in statistics between the fit with $T_{\rm PGo}=7.4$~keV and the fit with $T_{\rm PGo}=9.9$~keV is $\Delta{\rm C-stat}=+4.99$, which implies that a model with a temperature $T_{\rm PGo}=7.4$~keV is rejected in favor of a model with $T_{\rm PGo}=9.9$~keV at a confidence level of $97.5\%$. This suggests that the inner edge is in fact associated with a cold front, rather than with a shock front.


\section{Discussion and Conclusions}
\label{sec:summary}

\begin{figure*}
    \includegraphics[height=7.21cm]{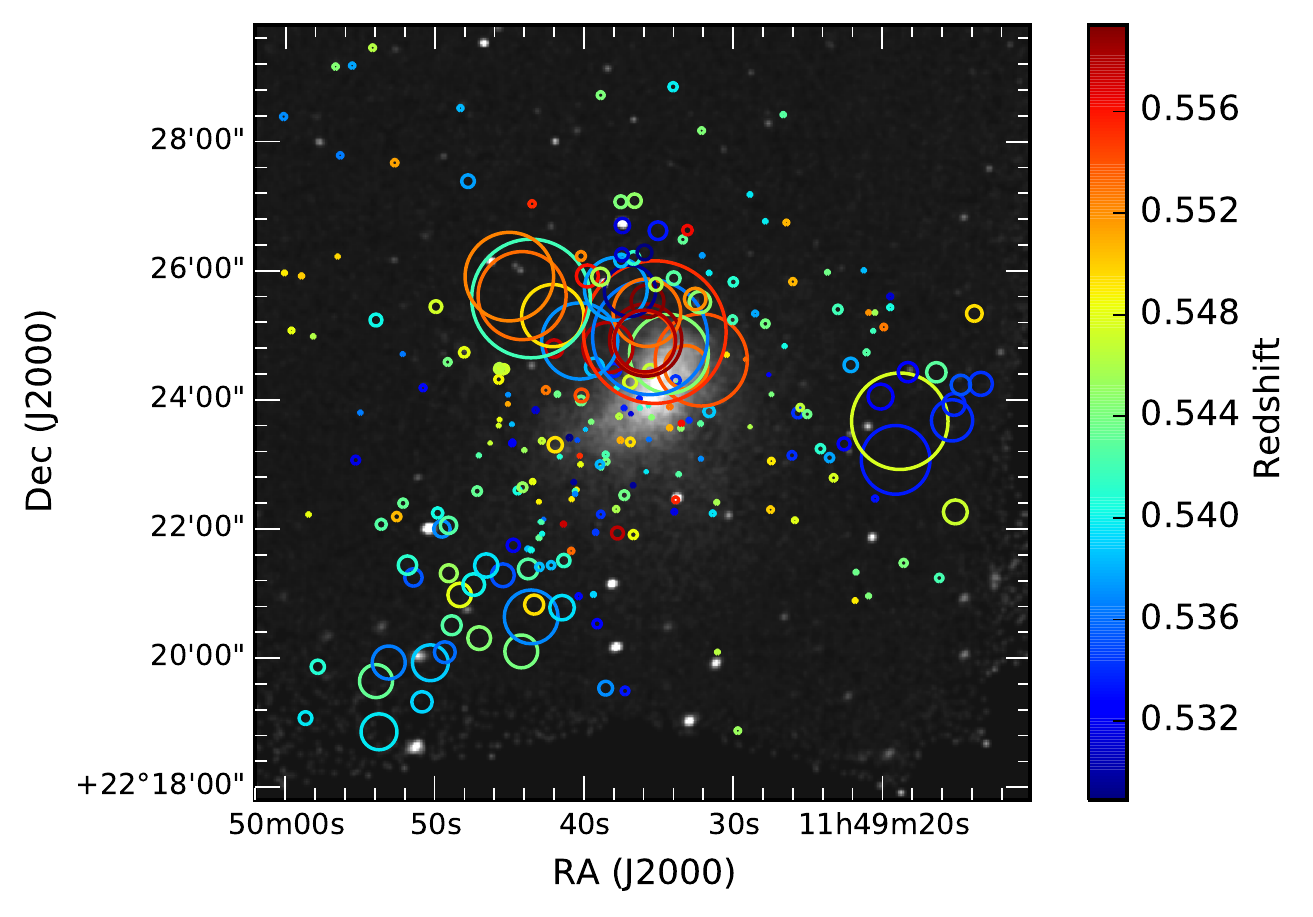}
    \includegraphics[height=7.21cm]{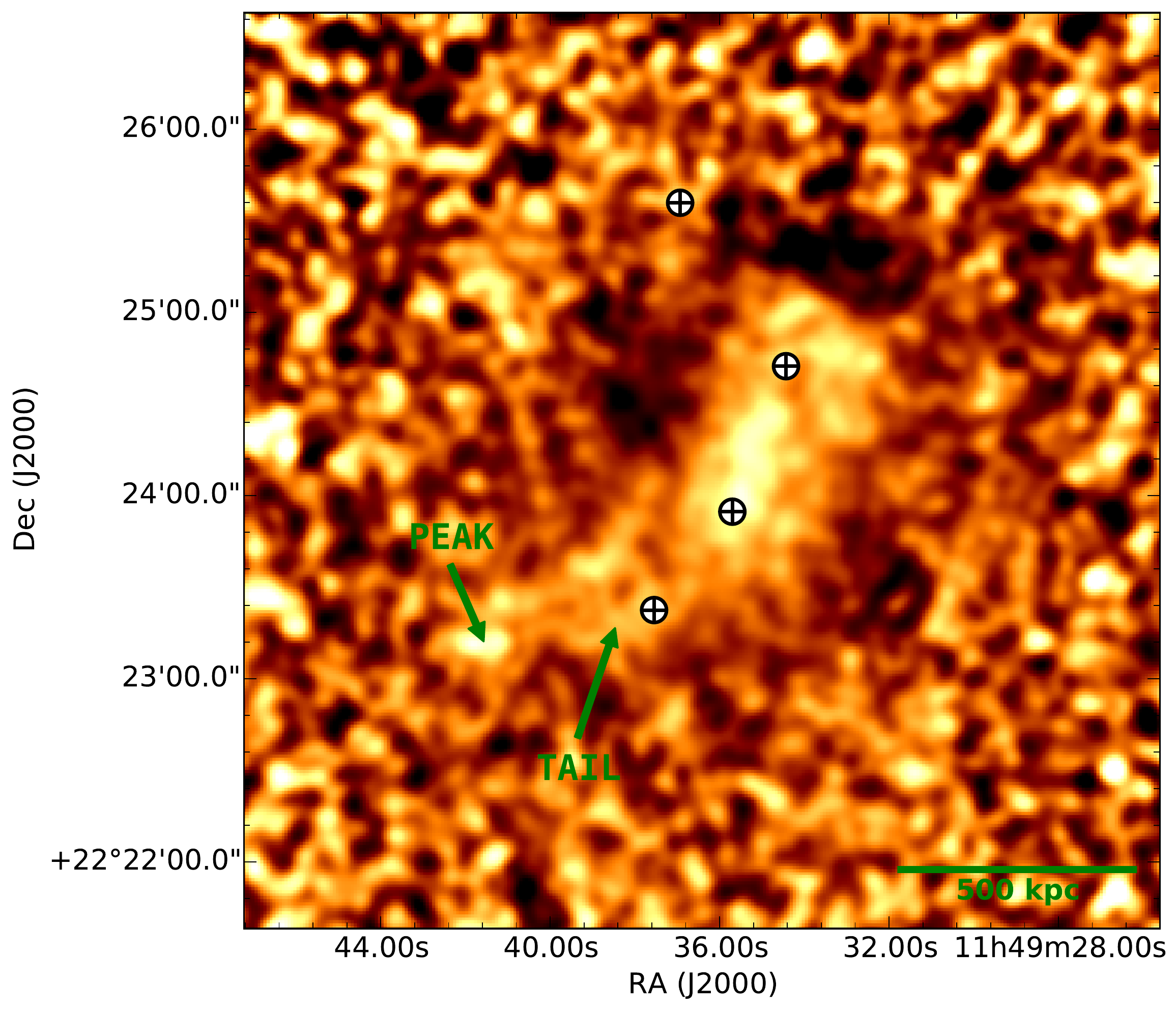}\vspace{0.15cm}
    \includegraphics[height=7.21cm]{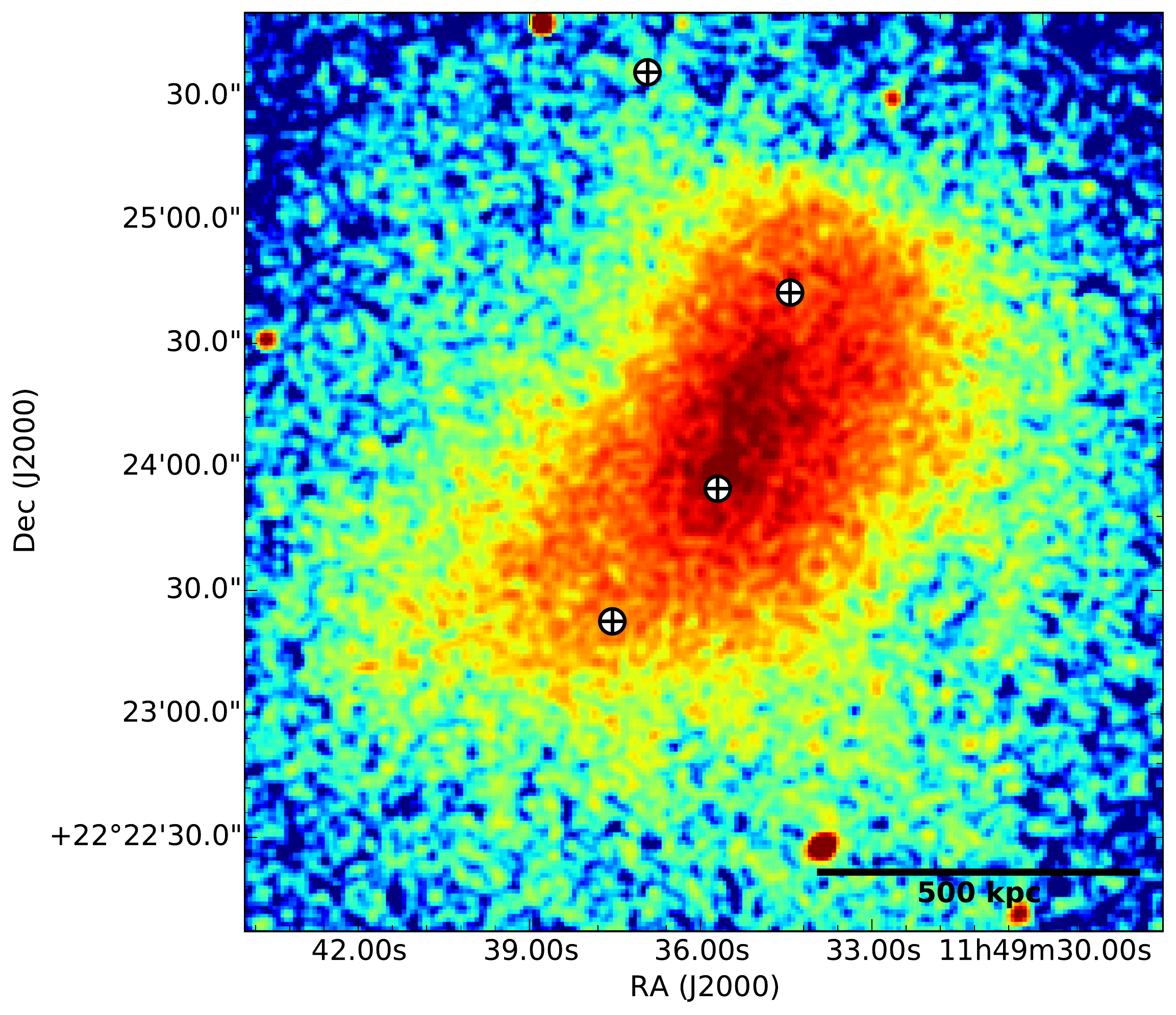} \hfill
    \includegraphics[height=7.21cm]{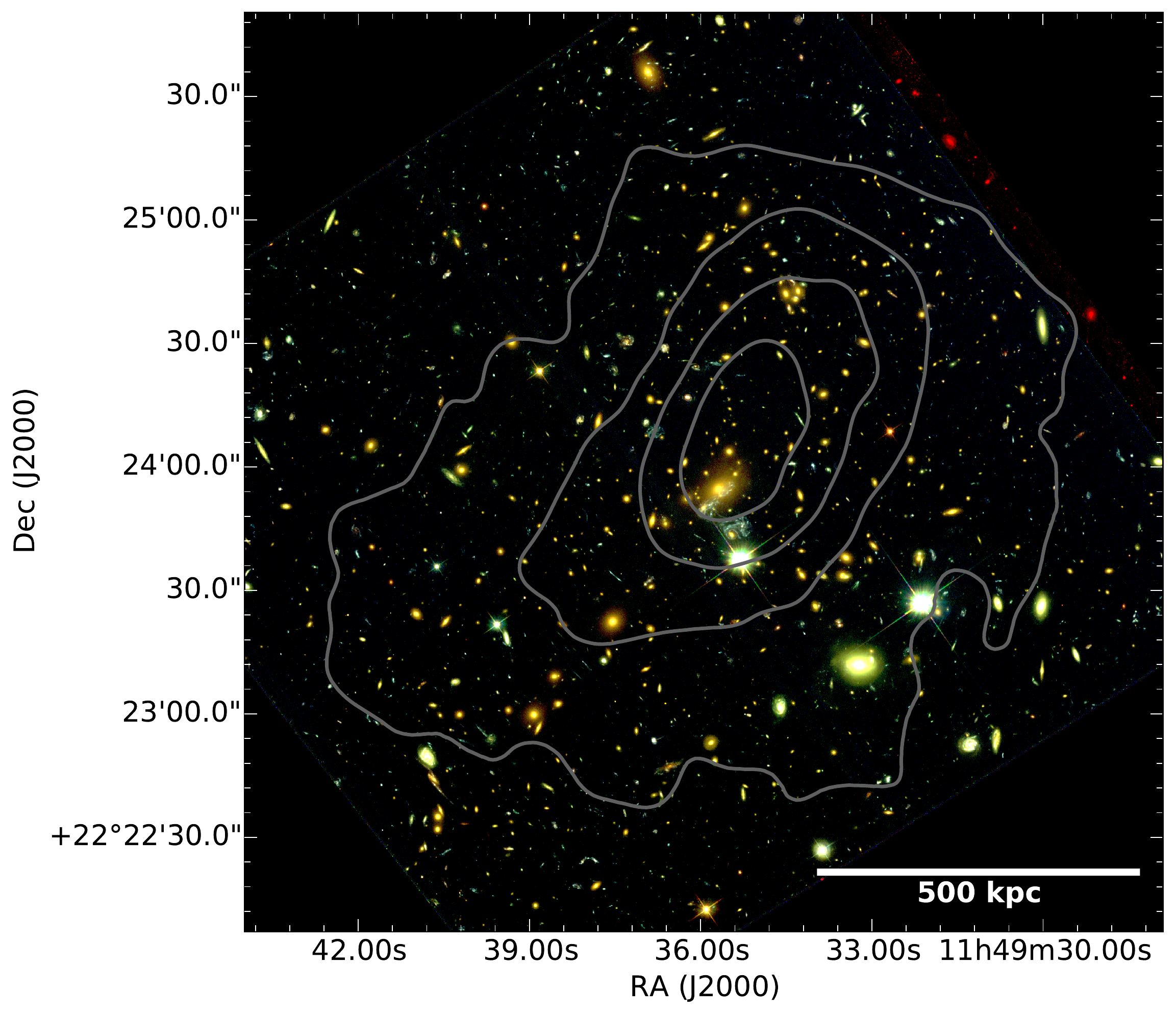}
    \caption{\emph{Top left:} DS map of MACS~J1149.6+2223. Circles show the projected locations of the cluster members, color coded for their redshift. The radii of the circles are proportional to the likelihood that the galaxy belongs to a substructure with a different line-of-sight velocity and/or a different velocity dispersion than the average. The \chandra\ surface brightness is shown in grayscale. \emph{Top right:} Unsharp masked image of MACS~J1149.6+2223. The image was created from two $0.5-4$~keV surface brightness maps smoothed with Gaussians of widths $3$ and $15$~arcsec, by dividing their difference to the map with the larger smoothing. The surface brightness maps had the point sources removed, and the ``gaps'' filled. Possible surface brightness edges on scales of $3-15$~arcsec ($20-95$~kpc) appear as strong gradients in the image. Putative edges are seen N-NE of the cluster center. The centers of the four DM halos identified by \citet{Smith2009} are shown with sun crosses. The tail emphasized by the unsharp masking and the region in which the tail's X-ray brightness peaks are indicated with arrows. \emph{Bottom left:} Same \chandra\ surface brightness map as in Figure~\ref{fig:sxmap}. The centers of the four DM halos identified by \citet{Smith2009} are shown with sun crosses. \emph{Bottom right:} HST RGB image of MACS~J1149.6+2223, based on observations using the F435W (blue), F606W (green), and F814W filters (red). { \chandra\ surface brightness contours are overlaid.} Both bottom panels show the same FOV, approximately $1.4\times 1.4$~Mpc$^{2}$. \label{fig:ds} \vspace{0.3cm}}
\end{figure*}

Our \chandra\ analysis of the X-ray properties of MACS~J1149.6+2223 revealed a previously unknown surface brightness edge located $\sim 350$~kpc N-NE of the cluster center. The edge is best fitted by a broken power-law model with a density compression $C=1.57_{-0.12}^{+0.09}$, and there is evidence that it is associated with a cold front; a shock front corresponding to a density compression of $1.57_{-0.12}^{+0.09}$ was rejected at a confidence level of $97.5\%$ based on the temperatures on both sides of the density jump. If confirmed, the cold front in MACS~J1149.6+2223, at $z=0.544$, is the most distant cold front discovered to date. Numerical simulations of mass-limited cluster samples have shown that the fraction of cold fronts up to redshift $z=1$ depends only weakly on redshift \citep{Hallman2010}. Observationally, the fraction of clusters with cold fronts has been estimated to be $\sim 50-70\%$ with \xmm\ and \chandra\ \citep{Markevitch2003,Ghizzardi2010}. The dearth of cold fronts detected at $z\sim 0.5$ through density and temperature jumps is most likely an observational limitation; detecting high-redshift cold fronts requires very deep observations with X-ray satellites that have arcsec spatial resolution. 

MACS~J1149.6+2223 is a major merger, and we find no evidence of a compact cool core. Therefore, the cold front is most likely a merger cold front, caused by a remnant subcluster core that decoupled from its dark matter halo during the merger event. In this scenario, the cold front should also have an associated bow shock . The putative surface brightness edge detected at $90\%$ confidence further out in the NE, $\sim 1$~Mpc away from the cluster center, could be this bow shock. {Approximating the cold front to be spherical, with a sphere radius of $\sim 350$~kpc, the distance between the cold front and the putative bow shock implies a Mach number of $1.3-1.4$ \citep[e.g.,][]{Farris1994}. Such a Mach number would be similar to those of the shocks detected in other galaxy clusters \citep[e.g.,][]{Macario2011,Russell2012}. No diffuse radio emission was detected near the location of the putative shock front. If the shock is confirmed, then the lack of radio emission could be explained by a low Mach number that would not allow it to efficiently accelerate particles. Most shock fronts are traced by radio relics, but there are also exceptions, such as in Abell 2146, in which no diffuse radio emission was detected near the shocks \citep{Russell2011}.}

The substructures in MACS~J1149.6+2223 were studied using the Dressler-Shectman (DS) test \citep{Dressler1988}. The resulting map is presented in Figure~\ref{fig:ds}. N-NE of the cluster, the DS map reveals substructure with a radial velocity component of $\sim 2500$~km~s$^{-1}$ relative to the average cluster redshift (Golovich et al., in preparation). The cold front and putative shock front detected N-NE of the center of MACS~J1149.6+2223 could have been caused by the collision of this high-velocity substructure with the main cluster. { However, as further discussed by Golovich et al. (in preparation), the collision that led to the formation of the cold front is not necessarily the same collision as the one that triggered the formation of the radio halo detected by \citet{Bonafede2012}.}

The DS map also shows multiple other substructures, which indicate that MACS~J1149.6+2223 is a very complex merger between more than only two subclusters. The complex nature of the merger is supported by its dark matter distribution \citep{Smith2009}, which is best described by a mass model consisting of four dark matter halos. In the bottom left panel of Figure~\ref{fig:ds}, we mark the location of the four dark matter halos on the \chandra\ surface brightness map. Given the high radial velocity differences between substructures seen in the DS map (top left panel in Figure~\ref{fig:ds} and Golovich et al., in preparation), at least some of the collisions must have a signficant line-of-sight component. The \chandra\ observations of the cluster support a complex merger scenario, in which the merger is seen primarily not in the plane of the sky. With the exception of the cold front and the putative shock front N-NE of the cluster center, we found no other surface brightness edges or temperature substructures. Throughout the ICM, the temperatures are very high, $\gtrsim 8$~keV. 

Furthermore, in the X-ray surface brightness map, one cannot visually separate the individual merging gas components, as can sometimes be done in simpler mergers occurring close to the plane of the sky (e.g., 1E~0657-56, CIZA~J2242.8+5301). In the top right panel of Figure~\ref{fig:ds}, we show the centers of the four dark matter halos identified by \citet{Smith2009} overlaid on the \chandra\ unsharp masked image. The unsharp masked image reveals a ``tail'' of gas with a length $\sim 500$~kpc, extending SE from the cluster center. The surface brightness of the tail peaks at the SE tip of the tail. The peak is due to two small regions of X-ray excess, each approximately 40~kpc in diameter. These regions are not clearly associated with X-ray or optical point sources, and they are offset $\sim 350$~kpc east from the southernmost dark matter peak. The origin of the dense gas causing the X-ray excess is unclear.

Given that MACS~J1149.6+2223 possibly consists of at least four merging DM halos, the lack of significant substructure in the ICM is somewhat surprising. For comparison, another Frontier Fields Cluster, MACS~J0717.5+3745, which also consists of at least four merging DM halos, has an ICM that appears significantly more disturbed and presents several X-ray features associated with shocks, cold fronts, and stripped gas \citep[][and van Weeren et al. 2015b, submitted]{vanWeeren2015}. The ICM of MACS~J1149.6+2223 could appear relatively regular if the merging subclusters collided perpendicularly to the plane of the sky. However, the centers of the DM halos identified by \citet{Smith2009} are separated, in projection, by $250-750$~kpc; these distances seem too large for the collisions to have occurred perpendicularly to the plane of the sky. Another possibility is that MACS~J1149.6+2223 is an old merger (see Golovich et al., in preparation, for a discussion of the merger scenario). If this is the case, then the collisionless DM halos are still in the process of merging, while the collisional gas halos have already merged. 

The faint radio halo discovered by \citet{Bonafede2012} could have a steep spectrum \citep[relatively shallow radio data suggests $\alpha \lesssim -2$;][]{Bonafede2012}. A steep spectrum could point to an old radio halo, in which particle aging caused a steepening of the spectrum, or to a less energetic merger \citep{Brunetti2008}. MACS~J1149.6+2223 is one of the most massive known clusters, so it seems unlikely that the merger is less energetic. An old radio halo, which could be seen in an old merger, would be consistent with the relatively regular ICM morphology. {While the radio halo could have been formed by an older collision between two of the subclusters in MACS~J1149.6+2223, a more recent collision between a different pair of subclusters might have given rise to the cold front seen $\sim 350$~kpc N-NE of the cluster center (Golovich et al., in preparation). The same collision that triggered the formation of the radio halo could also have resulted in the formation of the radio relic(s) detected by \citet{Bonafede2012}.}

By combining high-quality X-ray, radio, and optical/lensing observations of MACS~J1149.6+2223, one could potentially set constraints on its complex merger scenario. However, pinning down the merger geometry would be difficult.



\acknowledgments

We thank the referee for constructive comments.

GAO acknowledges support by NASA through a Hubble Fellowship grant HST-HF2-51345.001-A awarded by the Space Telescope Science Institute, which is operated by the Association of Universities for Research in Astronomy, Incorporated, under NASA contract NAS5-26555. RJvW is supported by NASA through the Einstein Postdoctoral grant number PF2-130104 awarded by the Chandra X-ray Center, which is operated by the Smithsonian Astrophysical Observatory for NASA, under contract NAS8-03060. AZ acknowledges support by NASA through a Hubble Fellowship grant HST-HF2-51334.001-A awarded by the Space Telescope Science Institute, which is operated by the Association of Universities for Research in Astronomy, Incorporated, under NASA contract NAS5-26555. This research was performed while TM held a National Research Council Research Associateship Award at the Naval Research Laboratory (NRL). FAS acknowledges support from Chandra grant G03-14131X. Part of this work performed under the auspices of the U.S. DOE by LLNL under Contract DE-AC52-07NA27344.

This research made use of NASA's Astrophysics Data System. This research also made use of the NASA/IPAC Extragalactic Database (NED), which is operated by the Jet Propulsion Laboratory, California Institute of Technology, under contract with the National Aeronautics and Space Administration. This research made use of APLpy, an open-source plotting package for Python hosted at \url{http://aplpy.github.com}. Some of the cosmological parameters in this paper were calculated using Ned Wright's cosmology calculator \citep{2006PASP..118.1711W}.

The scientific results reported in this article are based on observations made by the \chandra\ X-ray Observatory.



{\it Facilities:} \facility{Chandra (ACIS)}.




\bibliography{macsj1149-chandra}
\clearpage




\end{document}